\documentclass[twocolumn]{aastex62}

\newcommand{\arcs}{$^{\prime\prime}$}
\shorttitle{Sky-Limited IR Photometry with Warm InGaAs}
\shortauthors{Simcoe et al.}

\begin{document}

\title{Background-Limited Imaging in the Near-Infrared with Warm InGaAs
  Sensors: \\ Applications for Time-Domain Astronomy}

\correspondingauthor{Robert A. Simcoe}
\email{simcoe@space.mit.edu}

\author{Robert A. Simcoe}
\affiliation{MIT-Kavli Institute for Astrophysics and Space Research;
77 Massachusetts Ave., Cambridge, MA 02139}

\author{G\'{a}bor F\H{u}r\'{e}sz}
\affiliation{MIT-Kavli Institute for Astrophysics and Space Research;
77 Massachusetts Ave., Cambridge, MA 02139}

\author{Peter W. Sullivan}
\affiliation{Jet Propulsion Laboratory; 4800 Oak Grove Dr., Pasadena, CA 91109}

\author{Tim Hellickson}
\affiliation{MIT-Kavli Institute for Astrophysics and Space Research;
77 Massachusetts Ave., Cambridge, MA 02139}

\author{Andrew Malonis}
\affiliation{MIT-Kavli Institute for Astrophysics and Space Research;
77 Massachusetts Ave., Cambridge, MA 02139}

\author{Mansi Kasliwal}
\affiliation{Department of Astronomy, California Institute of Technology; 1216 California Blvd., Pasadena, CA 91125}

\author{Stephen A. Shectman}
\affiliation{Observatories of the Carnegie Institution for Science; 813 Santa Barbara St., PAsadena, CA 91101}

\author{Juna A. Kollmeier}
\affiliation{Observatories of the Carnegie Institution for Science; 813 Santa Barbara St., PAsadena, CA 91101}

\author{Anna Moore}
\affiliation{Research School of Astronomy and Astrophysics, Australian National University; \\ Mt. Stromlo Observatory, Weston Creek ACT2611, Australia}

\begin{abstract}

We describe test observations made with a customized $640\times512$
pixel Indium Gallium Arsenide (InGaAs) prototype astronomical camera
on the 100\arcsec ~DuPont telescope.  This is the first test of InGaAs
as a cost-effective alternative to HgCdTe for research-grade
astronomical observations.  The camera exhibits an instrument
background of 113 e-/sec/pixel (dark + thermal) at an operating
temperature of -40C for the sensor, maintained by a simple
thermo-electric cooler.  The optical train and mechanical structure
float at ambient temperature with no cold stop, in contrast to most IR
instruments which must be cooled to mitigate thermal backgrounds.
Measurements of the night sky using a reimager with plate scale of
$0.4$\arcsec/pixel show that the sky flux in $Y$ is comparable to the
dark current.  At $J$ the sky brightness exceeds dark current by a
factor of four, and hence dominates the noise budget.  The sensor read
noise of $\sim 43e^-$ falls below sky+dark noise for exposures of
$t>7$ seconds in $Y$ and 3.5 seconds in $J$.  We present test
observations of several selected science targets, including
high-significance detections of a lensed Type Ia supernova, a type IIb
supernova, and a $z=6.3$ quasar.  Deeper images are obtained for two
local galaxies monitored for IR transients, and a galaxy cluster at
$z=0.87$.  Finally, we observe a partial transit of the hot Jupiter
HATS34b, demonstrating the photometric stability required over several
hours to detect a $1.2\%$ transit depth at high significance.  A
tiling of available larger-format sensors would produce an IR survey
instrument with significant cost savings relative to HgCdTe-based
cameras, if one is willing to forego the $K$ band.  Such a camera
would be sensitive for a week or more to isotropic emission from
$r$-process kilonova ejecta similar to that observed in GW170817, over
the full 190 Mpc horizon of Advanced LIGO's design sensitivity for neutron
star mergers.

\end{abstract}

\section{Introduction}\label{sec_intro}

\begin{figure*} [t]
  \plotone{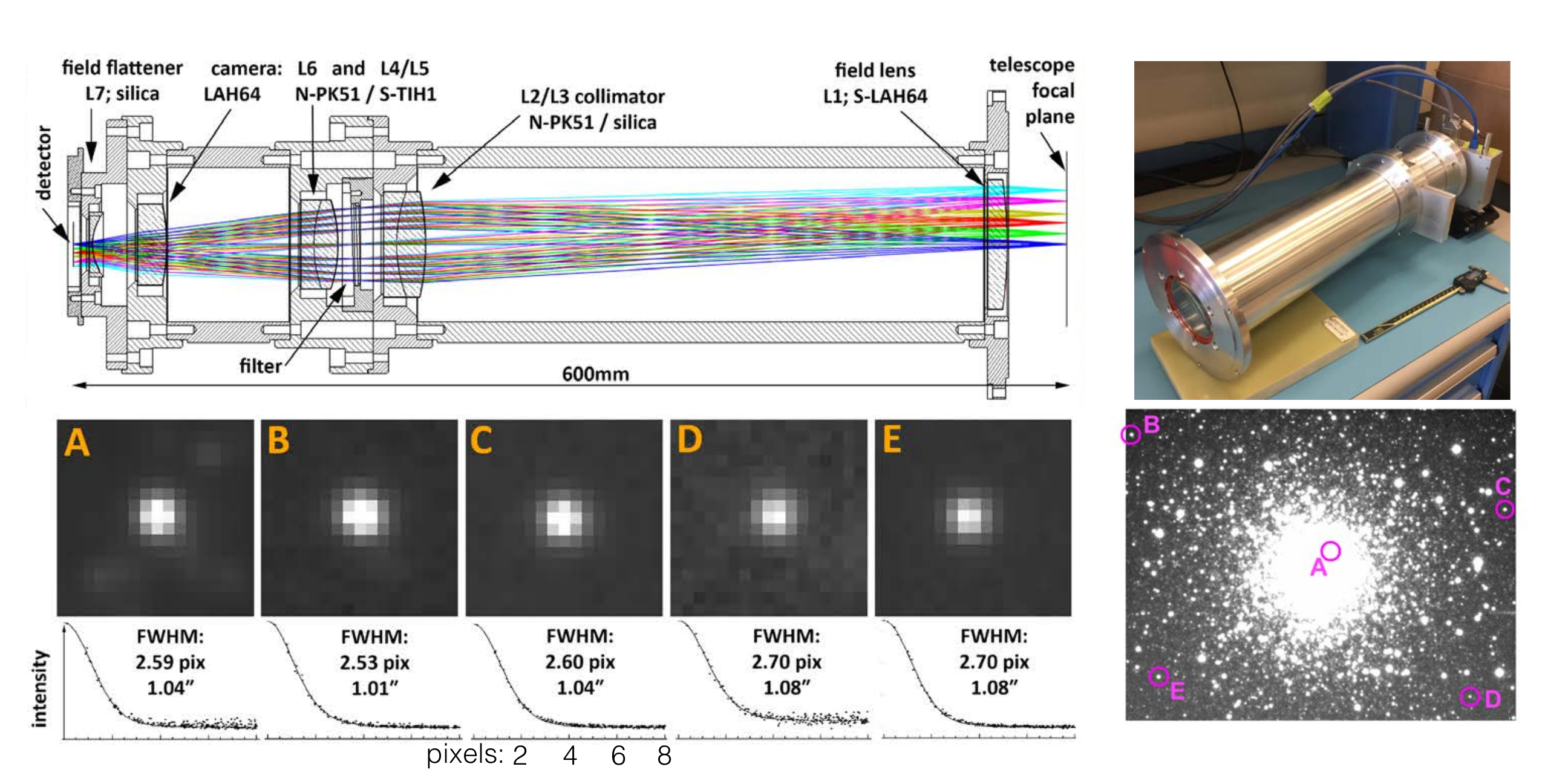}
  \caption{\label{fig:camera}Photograph of the InGaAs reimaging camera
    assembly (upper right) and cross-sectional view of the lens
    configuration (upper left).  The DuPont telescope focal plane is
    to the right of the field lens, and the sensor is located at left.
    The camera requires only boas voltages, separate power for the
    backing TEC, and USB3 (which is converted and transmitted over
    fiberoptic lines to the control room). Bottom row indicates PSF
    image quality across the field, measured using the globular
    cluster shown at bottom right for indicated field positions.}
\end{figure*} 

The high cost of infrared sensors has limited the size of focal plane
arrays for infrared sky surveys, despite the demand for this
capability to complement optical surveys now underway or planned
\citep{iptf,ztf,lsst}.  InGaAs focal planes offer a lower-cost
alternative to heritage designs based on HgCdTe, and sensors are
available with $15\mu$m pixels in up to 2k$\times$1k format.

The red cutoff for standard commercial InGaAs material is at
$\sim1.65\mu$m. Cameras made with these sensors only operate in the
$Y, J$, and short $H$ bands.  Owing to the lack of $K$-band
sensitivity, these cameras can be operated at higher temperature thus
relaxing requirements on sensor and instrument cooling. In addition,
InGaAs has lower dark current than HgCdTe at fixed temperature
\citep{beletic}, further facilitating warm operation.

Because InGaAs sensors are typically designed for high-background and
video applications, most commercially available cameras are not
suitable for astronomy research.  The high dark current from ambient
temperature operation, coupled with read noise measured in hundreds of
e- (RMS) compromises on-sky performance.  Some vendors are pushing the
noise envelope by either targeting low dark current from cooling, or
low read noise, but commercial cameras presently do not feature both
low dark current and low read noise simultaneously in a large format.
Specifically, none of the commercial cameras utilize non-destructive
reads to take advantage of Fowler \citep{fowler} or up-the-ramp
sampling \citep{chapman1990} as is common in astronomical
research-grade cameras.

Recently, progress in material growth has yielded sensors with lower
dark current that are packaged with integrated thermoelectric coolers
(TECs) to lower the noise floor.  To miticate read noise, we have
built a camera with a 640$\times$512 pixel InGaAs sensor, and
implemented non-destructive reads via a custom daughter board and FPGA
to mitigate read noise.  Our laboratory measurements indicate that
this system should achieve sky-background limited noise performance on
a 1-meter telescope with 1\arcsec ~pixels, or larger telescopes with
smaller pixels in the $J$ and $H$ bands, and marginally for $Y$ as
well.

This report describes on-sky performance tests of this prototype
camera on the 2.5-meter DuPont telescope at Las Campanas Observatory,
Chile.  During the course of a 3-night engineering run during November
2016 bright time, we mounted the camera with a small set of reimaging
optics to measure broadband sky backgrounds, obtain photometric zero
points, test imaging depth and photometric stability.  We observed
several IR transients as well as faint static sky targets to
demonstrate possible science applications of an imager in this
configuration.  Based on experience with this system we project the
performance of warm InGaAs imagers for selected science appliations.

\section{Sensor Selection}\label{sec_overview}

All tests described here were made with a 640$\times$512 AP1121 sensor
from FLIR electro-optical components.  This is very similar to the
APS640C camera described in \citet{sullivan_ap640c}, but with 15$\mu$m
rather than $25\mu$m pixels\citep{sullivan_ap1121,sullivan_thesis}.
The dark signal continues to drop as the temperature is lowered, to
the $-45$C limit of tests with our apparatus.

The AP1121 has an electronic shutter and CTIA pixel architecture,
unlike the HAWAII family whose pixels have source-follower amplifiers
\citep{beletic}.  We run the AP1121 at high frame rate and average
multiple frames using sample-up-the-ramp mode to reduce read noise.

Primary cooling power for the sensor is derived from the
thermo-electric cooler (TEC) integrated into its package. The InGaAs
substrate and ROIC are protected by the package's vacuum seal, which
admits light through an AR-coated sapphire window.  In our camera
head, the warm side of the on-sensor TEC abuts a copper post, which
has a secondary high-capacity ``backing'' TEC on its opposite side,
powered by a dedicated supply.  The warm side of the backing TEC mates
to a standard CPU water cooling block, of the variety seen in many
gaming or overclocked PCs.  Standard 0.75-inch Buna-N hoses connect
the cooling block to a ThermoTek recirculating bath chiller.  We run
the bath with distilled water at +5C, which cools the TEC but did not
result in condensation on hoses or within the camera head.  To
mitigate condensation on the window of the sensor package, we run a
N$_2$ gas line into the camera head with a regulator to provide a
gentle positive flow of dry air.

This design does not maintain the sensor package in vacuum, nor does
it achieve deep cooling, since the TEC stack implementation (which
employs an long thermo-mechanical path) is very inefficient.  However
it did allow us to operate the camera in the range of $T=-40$ to
$-45$C, adequate for our testing purposes.

\section{Reimaging Camera and Filters}\label{sec_opticaldesign}

To resample the focal plane into a scientifically realistic
configuration, we assembled a modest reimaging camera that converts
the plate scale of the 100-inch DuPont telescope from its native value
of 0.16\arcsec ~per pixel to 0.4\arcsec.  This provides Nyquist
sampling for all but the best seeing conditions at Las Campanas, and
is set as coarse as is practical to mimic the field-of-view
requirements for a wide-field survey instrument.  The overall field of
view of the 640$\times512$ sensor is 4.3\arcmin$\times$3.4\arcmin.

The reimager optics (Figure \ref{fig:camera}) begin with a 72mm
diameter (clear aperture) plano-convex field lens, which circumscribes
the projected field of view on the telescope focal plane and defines a
pupil location inside the barrel near the filters.  The diverging beam
is collimated by a doublet of diameter 58mm, just prior to the optical
pupil.

Notably, this configuration has no cold Lyot stop---the
optics are completely warm and float at ambient temperature.  Such a
design could easily accommodate a stop if one were desired, but we
will demonstrate below that the $K$-blind nature of the InGaAs sensor
makes the instrument sky-noise dominated.

\begin{figure} [t]
  \plotone{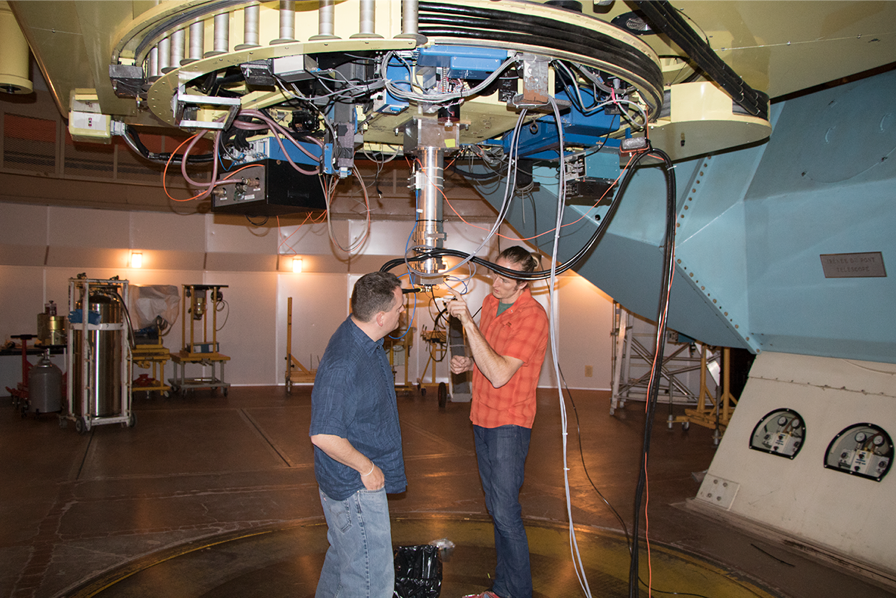}
  \caption{\small \label{fig:on_dupont} InGaAs prototype
    camera mounted on the DuPont telescope's Instrument Mounting Base.
    Buna-N coolant hoses and fiber-optic interface are draped at right
    from the port, the gray cables bring power for the sensor and
    thermo-electric backing coolers.}
\end{figure} 

Our MKO-$Y$ and $J$ band filters \citep{tokunaga} mount just beyond
the pupil in a detent-indexed slide, which tilts the filters by 2
degrees to avoid ghosting.  Because our sensor substrate is classical
InGaAs material and is not prepared for extended-blue sensitivity (due
to cost and availability) there is a slight throughput loss from QE
rollover at the bluest edge of the $Y$ band, but this has only a small
effect on overall sensitivity.  Although InGaAs is also photosensitive
over much of the $H$ band, we did not test in this region because our
main goal was to establish whether the camera would be
background-limited, and we were confident that it would be in $H$
because of the much brighter sky emission.

A 50 mm diameter doublet is used in conjunction with a plano-convex
singlet and field flattener to refocus the beam onto the sensor at the
desired pixel scale.  All lenses were treated with broad-band
anti-reflection coatings from Evaporated Coatings, Inc.

The lenses were bonded into precision-machined bezels using RTV60.
Each of the lens groups contained one plano surface to simplify axial
registration against datum surfaces on the bezels.  Laboratory test
images taken with a USAF target placed at the location of the DuPont
focal plane verified full MTF contrast at line spacings of 71$\mu$m
(0.77\arcsec) and partial contrast at 62$\mu$m (0.67\arcsec) for a
projected pixel scale of 37$\mu$m.  This indicates that image quality
is limited by the telescope and sensor commbination, and aberrations
from the camera contribute at the $<1$ pixel level.  Astrometric
calibration of images on the sky yield an as-built pixel scale of
0.396\arcsec (design is 0.4).

We shipped the camera to Las Campanas Observatory fully assembled, and
bolted to the standard Instrument Mounting Base of the DuPont so as to
use the observatory guiders during the night.  We placed the chiller
on the dome floor and draped cables from hard points on the telescope.
The cables included a USB3-to-fiber converter which ran to our control
laptop in the control room.  We located all power supplies in the
control room, with long cable runs to the instrument port, to
facilitate diagnostic monitoring and power cycles if needed.

\section{Detector Performance}\label{sec_detectors}

\subsection{Dark Current}\label{sec_dark}

\begin{figure*}[t]
  \plottwo{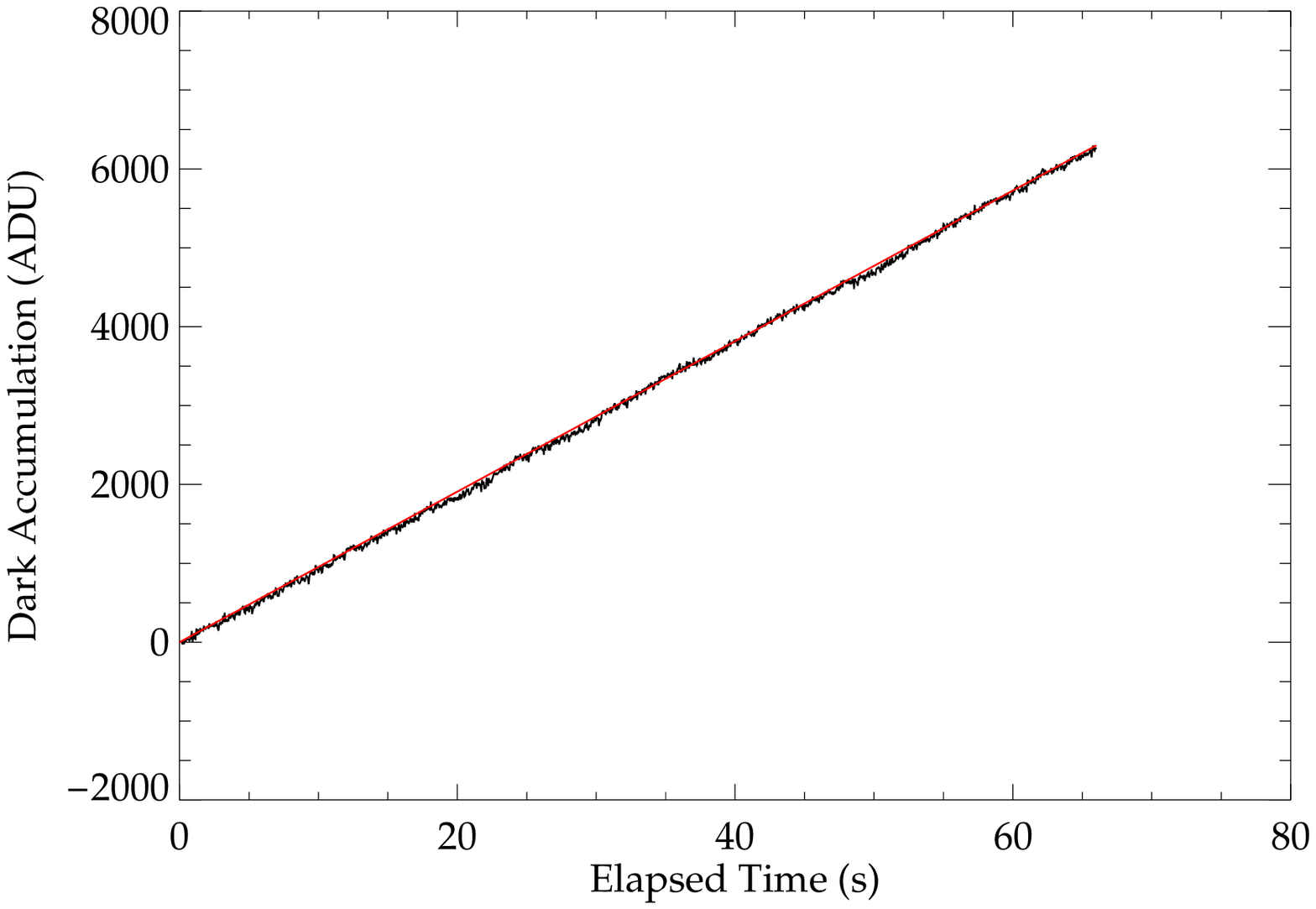}{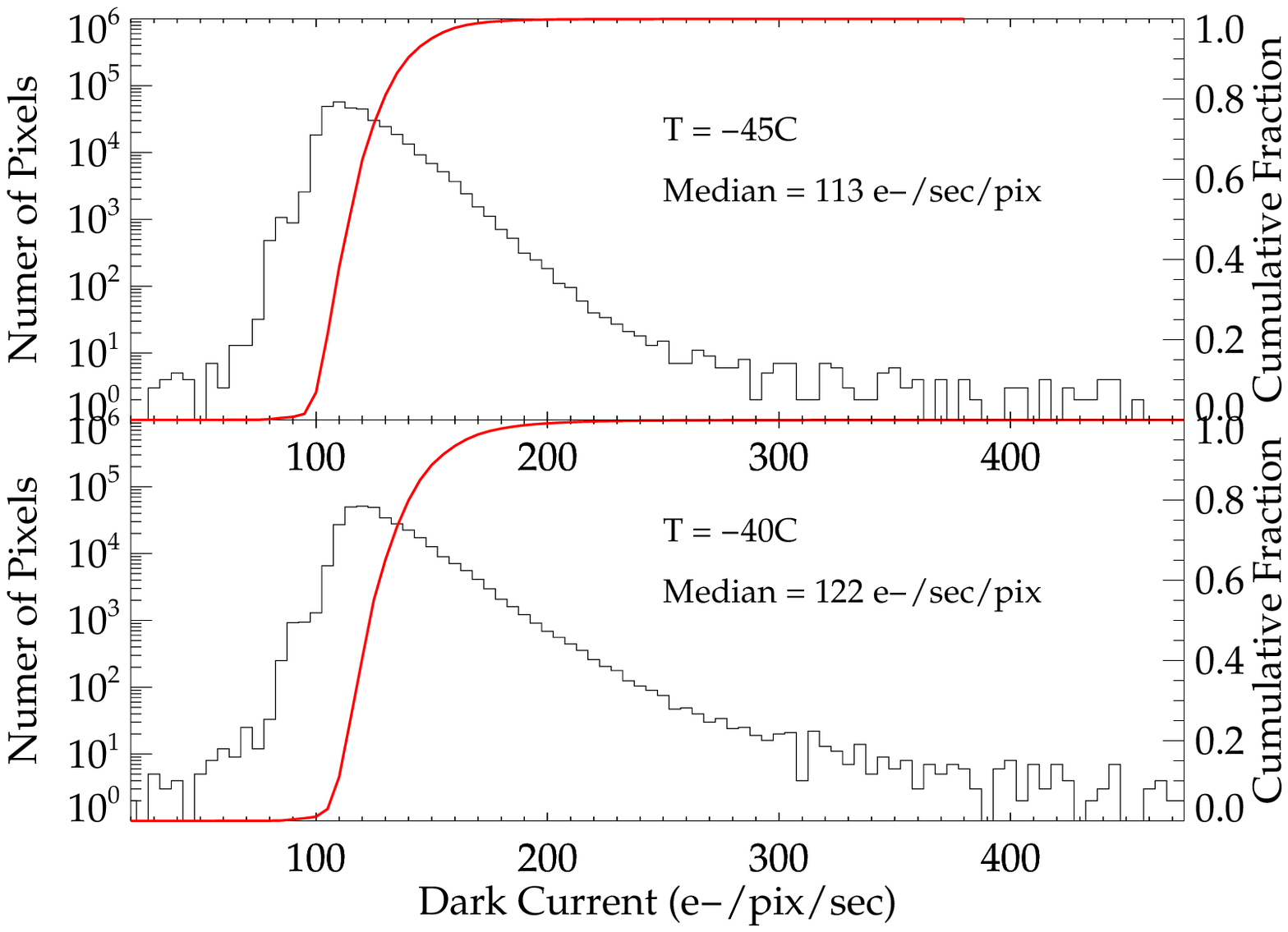}
  \caption{\label{fig:dark_measured}\small (Left) Example of a dark
    current ramp fit for a single pixel at $T=-45C$, accumulating at
    approximately 100 counts/second.  Black represents measured
    counts, with the linear fit shown as a red solid line.  (Right)
    Dark current measurements in e-/sec as mounted on the telescope. }
\end{figure*} 

We obtained measurements of the combined background from sensor dark
current and thermal emissivity by closing the dome and mirror covers
during nighttime hours and taking one-minute ramps with 1200 samples.
After correcting these frames for linearity, we performed a linear fit
of the dark count slope versus time for each pixel on the array.
Figure \ref{fig:dark_measured} shows an example of one such fit for a
single pixel in the left panel, and the right panel shows statistics
of the per-pixel dark current across the array, assuming a gain of
1.17 e-/ADU as measured previously in the lab \citep{sullivan_thesis}.

We operated our first two nights at $T=-40$C and the third night at
$T=-45$C, and measured the dark current at both temperatures.  The
median dark current at the lower temperature is 113 e-/sec/pixel.  A
high-end tail is visible and most prominent at the array corners; this
tail is suppressed at colder temperatures, and 90\% of pixels have a
dark value of $\le 140$ e-/sec as seen in the red cumulative
histogram.

This measured dark current is high relative to cryogenically cooled
HgCdTe. However at equivalent temperature ($T=-40$C) and pixel size,
commercial 1.6$\mu $m-cutoff HgCdTe would have a much higher dark
level of 100,000 e-/sec/pixel \citep{beletic}, and $2.5\mu m$ HgCdTe
would be higher yet. This is the fundamental factor enabling use of
InGaAs for lower cost warm instruments. Our prior laboratory
measurements indicate that further reduction in dark current is
possible at lower operating temperature, allowing one to tailor dark
current to different site conditions through straightforward thermal
engineering.

\subsection{Read Noise}\label{sec_readnoise}

We did not measure read noise separately at the telescope, but instead
used laboratory values taken prior to on-sky deployment. Using flat
fields taken at varying illumination levels below the full-well
(corrected for non-linearity), we calculated the conversion gain from
the slope at 1.17 e-/ADU by regressing signal variance against
mean. Extrapolating this fit to zero mean flux, we obtained a
single-sample read noise of 59 e- RMS.

The delivered read noise is reduced substantially through
non-destructive sampling. For up-the-ramp integrations of 64 samples
we measure read noise of $43e^-$ RMS; during on-sky operations we
always operated in SUTR mode, and except for bright standard stars our
ramps always exceed 16 reads, keeping read noise within this bound.
For comparable read noise values, Poisson noise from dark+sky counts
overtakes read noise within a 7 second exposure in $Y$ and 3.5 seconds
in $J$, using the sky brightness estimates measured as described
below.

\subsection{Linearity}\label{sec_linearity}

Non-linearity is a known issue with all IR sensors and our previous
measurements in the lab revealed non-linearity at the $\sim 3-5\%$
level, varying across the sensor.  We calibrate out nonlinearity using
flat field ramps run to saturation at the start of the observing run.
For each pixel, we fit a 4th order polynomial to the residual counts
relative to a straight linear fit, regressing the residuals against
count rate over the first 35,000 counts in each pixel.  We store the
polynomial coefficients for each individual pixel and correct each
science and calibration frame as the first step in data reduction
(before calculating the SUTR slope).  

\begin{figure} [t]
  \plotone{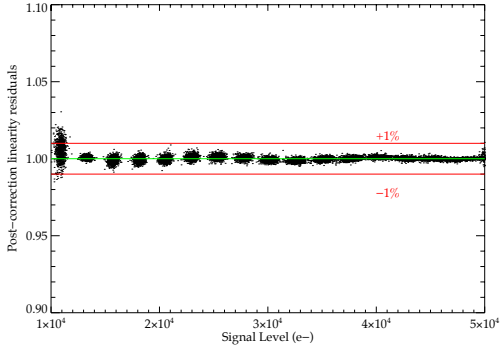}
  \caption{\label{fig:nonlin}Residuals of an individual flat field
    exposure relative to a straight linear fit, after performing a
    fourth-order polynomial nonlinearity correction. Slightly higher
    residuals are measured at low count rate, but the sensor is
    essentially never run in this regime because of robust thermal and
    sky backgrounds. }
\end{figure}

This method reduces the residual observed non-linearity to $\sim
0.5\%$ or less (Figure \ref{fig:nonlin}).  There is evidence in the
residuals that a higher order polynominal could reduce nonlinearity
yet further, at the risk of over-fitting. We also observe a higher
non-linearity near the start of our ramps (i.e. 1\% at the first
sample), but the array is essentially never operated in this regime
because of robust dark current and sky backgrounds.

\subsection{Data Reduction}\label{sec_datareduction}

Data from the camera are stored as SUTR FITS arrays containing
individual reads from each sensor reset.  A small set of IDL routines
is used to correct for non-linearity in the counts and then calculate
the SUTR slope as described in \citet{BenfordLauerMott2008}.  The
output is expressed in terms of e-/second.  The non-linearity
correction is critical for maintaining uniformity of the background
counts, as the size of the correction varies coherently across the
array.

We then subtract a dark current frame (also in e-/second) compiled as
described above.  The remaining flux from the sky and sources is
normalized by a flat field composite constructed from stacked twilight
sky exposures taken with the DuPont.  The flat field reference is
corrected for non-linearity and dark current, and normalized by the
median count rate to derive a unity-median calibration frame that is
divided into the science frame's count rate.  Individual pixels showed
gain variations of 1-5\% relative to the median value.

For fields requiring deep photometry, we combined multiple exposures
taken at different telescope dither positions. We used the {\tt fitsh}
processing software \citep{fitsh}, which matches and registers the
images to a common reference using direct photometry of (sometimes
faint) stars.

We subtracted a spatially constant background flux from each image and
applied an overall scaling to match fluxes between exposures.  The
final composite was constructed from a median of the registered,
background-subtracted, and flux-scaled frames.  No weights were
applied for the average since a median rather then mean was used for
the operation.  Finally, we reapplied approximate absolute astrometric
solutions generated by the Scamp and Swarp packages (from
astr0matic.net) to the final stacks for future convenience.

\section{On-sky Performance}\label{sec_performance}

The configured camera was used during three engineering nights on UT
2016 November 11, 12, and 13.

\subsection{Sky Backgrounds}\label{sec_skybg}

\begin{figure} [t]
  \plotone{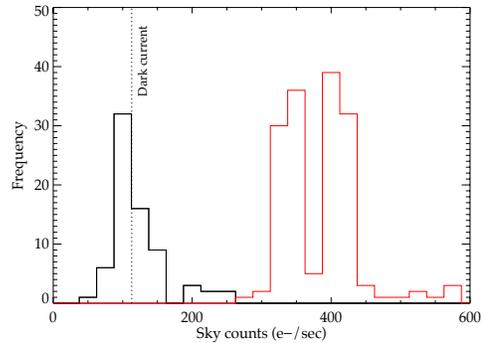}
  \caption{\small\label{fig:dark_sky}Histogram of median sky counts
    observed in individual frames taken on the DuPont.  The black
    histogram depicts the $Y$ band sky and red curve indicates $J$
    background.  Dotted line indicates the measured dark current at
    $T=-45$C.  $J$ band observations are well into the sky-limited
    regime, while Poisson noise from the dark current and sky
    contribute roughly equally to the total noise budget in $Y$.  This
    can be mitigated through colder operation or a coarser pixel
    scale.}
\end{figure} 

We gathered statistics on the sky backgrounds by recording the median
count rate for every exposure taken during the run, in each filter.  A
histogram of these values is shown in Figure \ref{fig:dark_sky}.

The Y band background is characterized by a unimodal distribution
centered around a median count rate of 103 counts/sec/pixel, or 121
$e^-$/sec/pixel.  This is nearly identical to the measured dark
current at $-40$C and slightly higher than the dark rate for $-45$C,
indicating that we are nearly sky-noise limited even the band with the
darkest sky.

\begin{figure*}[t]
  \plottwo{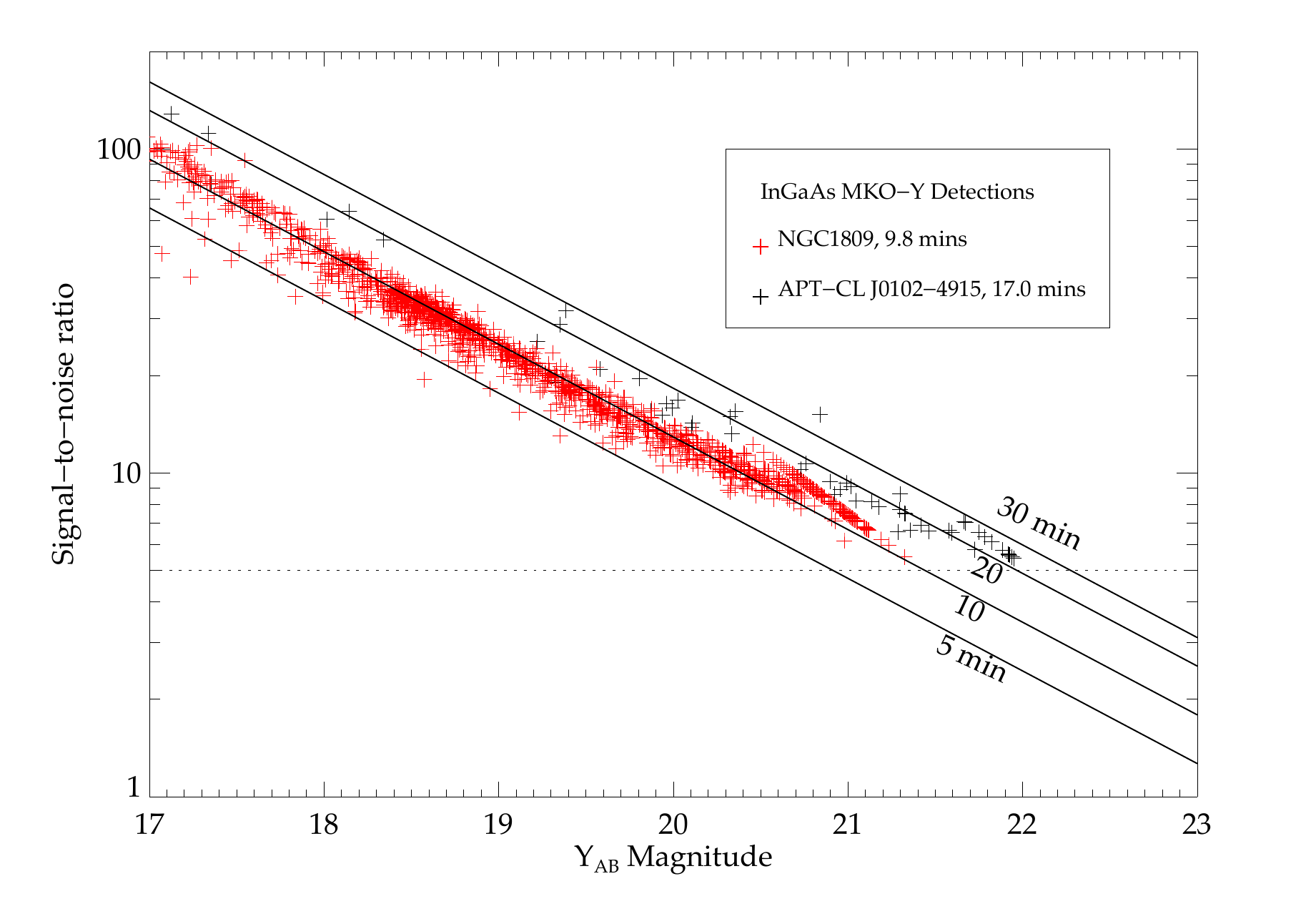}{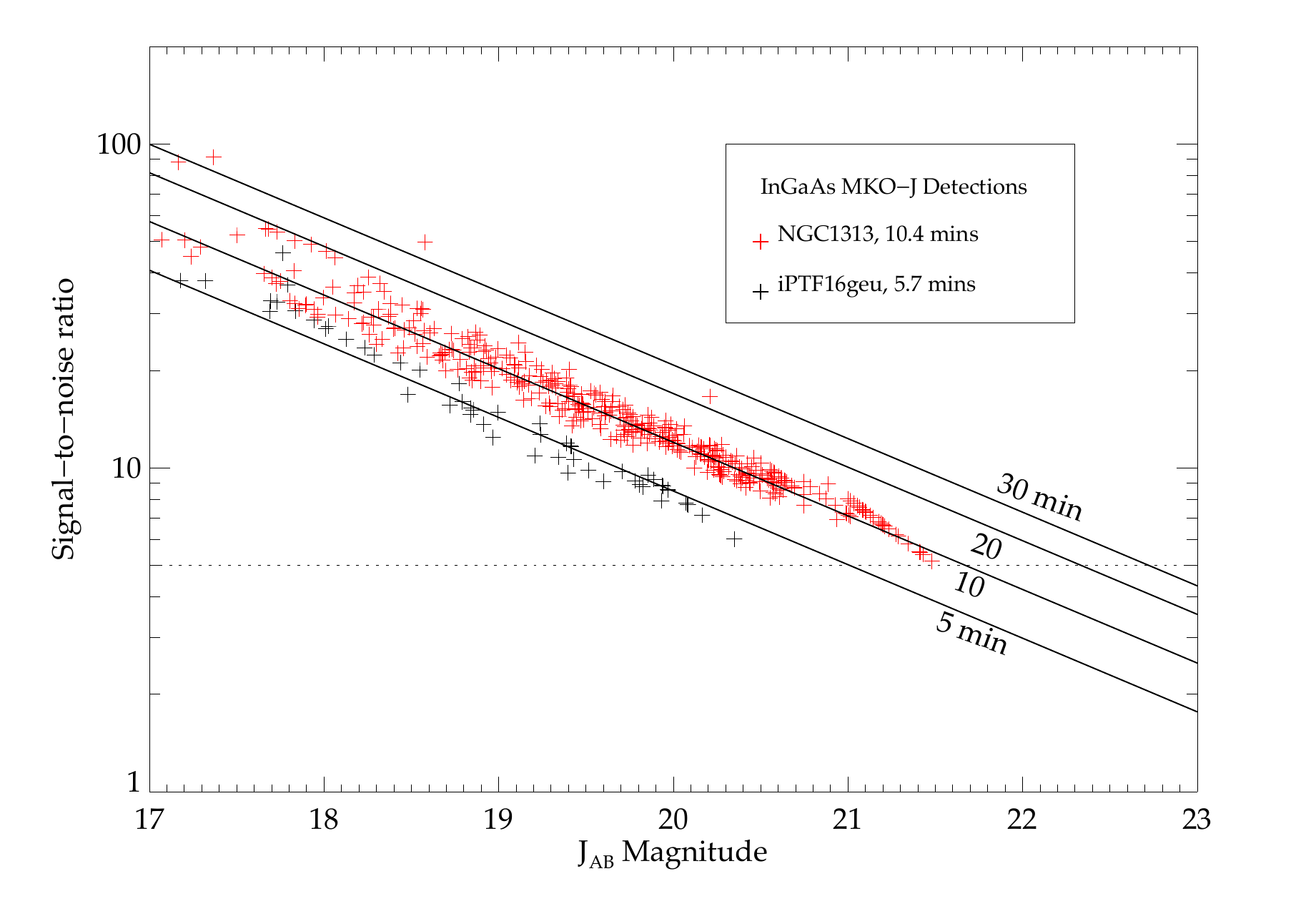}
  \caption{\label{fig:depth}\small Photometric depth,
    displayed as signal-to-noise ratio over apparent magnitude in
    three fields with different filters and exposure times.  The SNR
    is calculated from the isophotoal fluxes (and errors) of objects
    measured with Sextractor.}
\end{figure*}

\begin{deluxetable}{c c c}[b]
\tablewidth{0pc}
\tablecaption{On-Sky InGaAs Prototype Performance}
\tablehead{{Quantity}&{$Y$ value} & {$J$ value}}
\startdata
Zero point (1 e-/sec)   & 24.53 & 25.27 \\
Sky background (e-/sec/0.4\arcs ~pixel) & 120   & 413 \\
Sky background (mag/sq. arcsec) & 17.5 & 16.8 \\
Dark+thermal background (e-/sec/pixel) & 113 & 113 
\enddata
\end{deluxetable}

The $J$ band sky counts are bimodal, because some of our pointings
were quite close to the moon.  The median $J$ sky flux was 352
counts/sec/pixel, or 412 $e^-$/sec, a factor of $\sim 4$ higher than
the dark current.  Observations in $J$ (and by extension, $H$ as well)
will be dominated by Poisson noise from the sky, with some margin.

We estimated a calibration of the night sky brightness using
observations of the spectrophotometric standard star Feige 110 taken
during nights 2 and 3 of the run.  Using data from the CALSPEC archive
at STScI and measured transmission curves of our filters supplied by
vendors, we estimated the bandpass center of each filter and compared
with measured count rates to establish photometric zero points.  The
measured zero points and sky backgrounds are reported in Table 1.

From these calibrations, we estimated the median $Y$-band sky surface
brightness at 17.45 AB magnitudes per square arcsecond, which may be
compared with the measured value of 17.50 determined at the same site
for Magellan/FourStar
\citep{fourstar}\footnote{https://magellantech.obs.carnegiescience.edu/0sac/20110912/ FourStar\_Commissioning\_Report\_15aug2011.pdf}.
Likewise the $J$ band surface brightness is 16.84 AB magnitudes per
square arcsecond, compared to 16.90 measured by FourStar.

There remain systematic uncertaintes in our calibration from not
accounting the detailed sensitivity curve of the instrument or
variation in the SED of the standard star across each filter bandpass.
These calculations merely indicate that our zero points and
methodology produce consistent answers with other instruments at the
same site with established heritage.  Importantly, they also indicate
the possibility of achieving largely sky-limited noise performance
using a warm optical train with no Lyot stop, and a modestly-cooled
sensor.

If one wishes to obtain a more favorable noise budget, then further
sensor cooling should result in even lower dark current values
\citep{sullivan_thesis}.  Alternatively, adaptation of the optical
design to deliver 0.5\arcsec pixels rather than 0.4\arcsec would
increase the sky background by 50\% and cover a wider field.  This
must be traded against the desire to properly sample seeing and
achieve maximum point source sensitivity.

\subsection{Photometric Depth}\label{sec_sensitivity}

We compiled imaging data on multiple fields with differing depths as
we exercised the instrument in various configurations and settled on
observing strategies.  Because the camera is background limited, the
SNR should scale as $\sqrt{t}$, but a goal of the run was to establish
the baseline for this scaling.  

In Figure \ref{fig:depth} we show photometry for two fields apiece in
$Y$ and $J$.  The measurements are generated using Sextractor
\citep{sextractor} with a detection threshold of $5\sigma$ yielding
object catalogs of object isophotal AB magnitudes (uncorrected for
aperture), which we plot against SNR. A regression on the 10-minute
exposure photometry (for NGC1809 in $Y$ and NGC1313 in $J$, shown as
solid lines) confirms that sensitivity scales approximately as
$\sqrt{t_{exp}}$ as expected for other fields observed at different
depth.

The shallower slope of the $J$ band curves reflects the higher sky
background, but the overall sensitivity is similar because of the
sensor's higher QE in $J$, and correspondingly higher zero point.
Both filters reach a limiting magnitude of approximately $AB=21$ in 5
minutes, $AB=21.5$ in 10 minutes, and $AB=22-22.4$ in 20 minutes.

For our $J$ band observations, we cross-checked our photometry
calibrated with Feige 110 against 2MASS \citep{2mass} stars in the field
and adjusted the zero points to bring the two into agreement.
Typically these corrections were $\sim 0.05$ magnitudes except in one
case where we observed $0.15$ magnitudes of extinction from clouds.
For $Y$ we did not have a photometric reference catalog covering
observed fields and used our Feige110 calibrations without correction.

\begin{figure*}[!t]
  \center{\plottwo{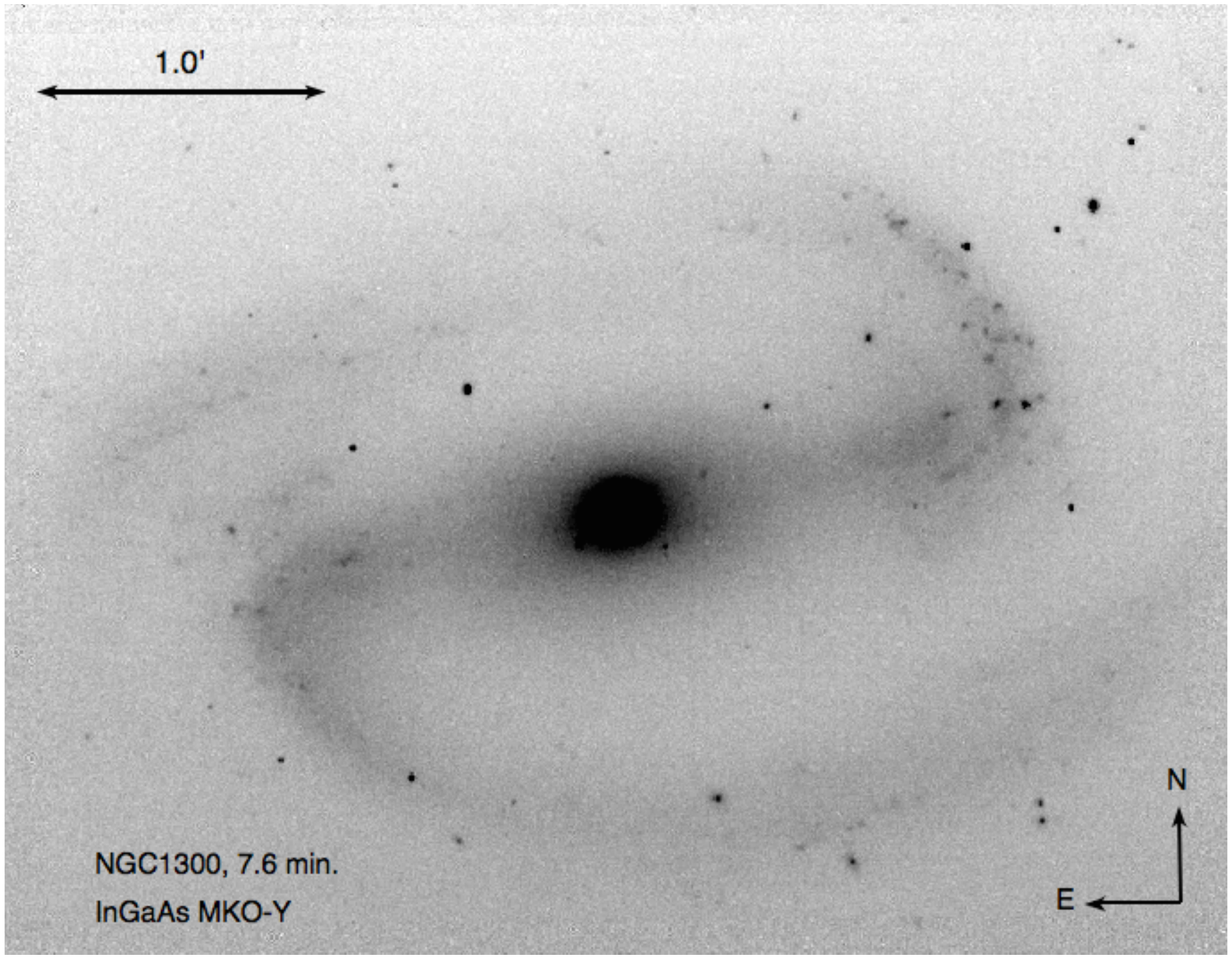}{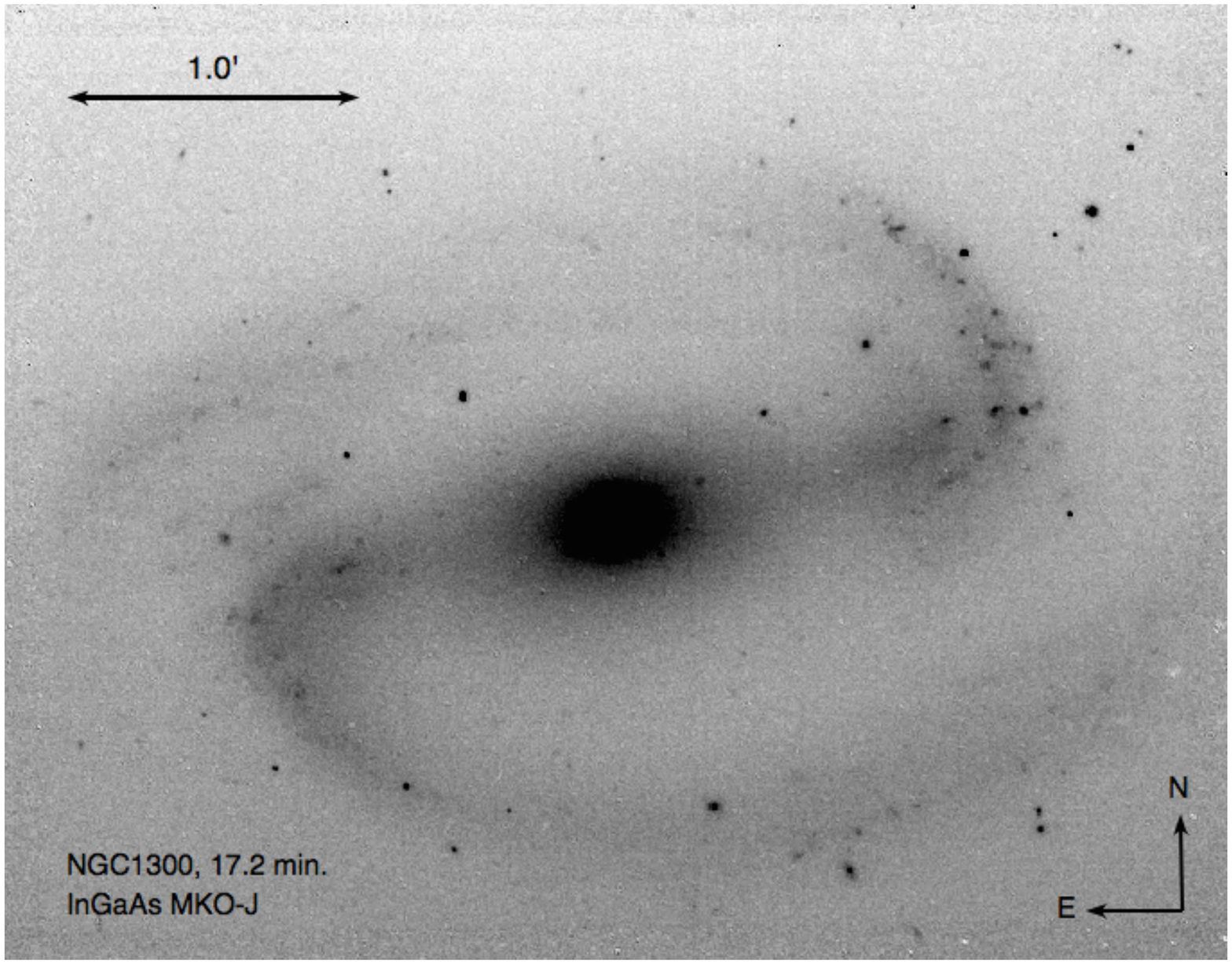}}
  \center{\plottwo{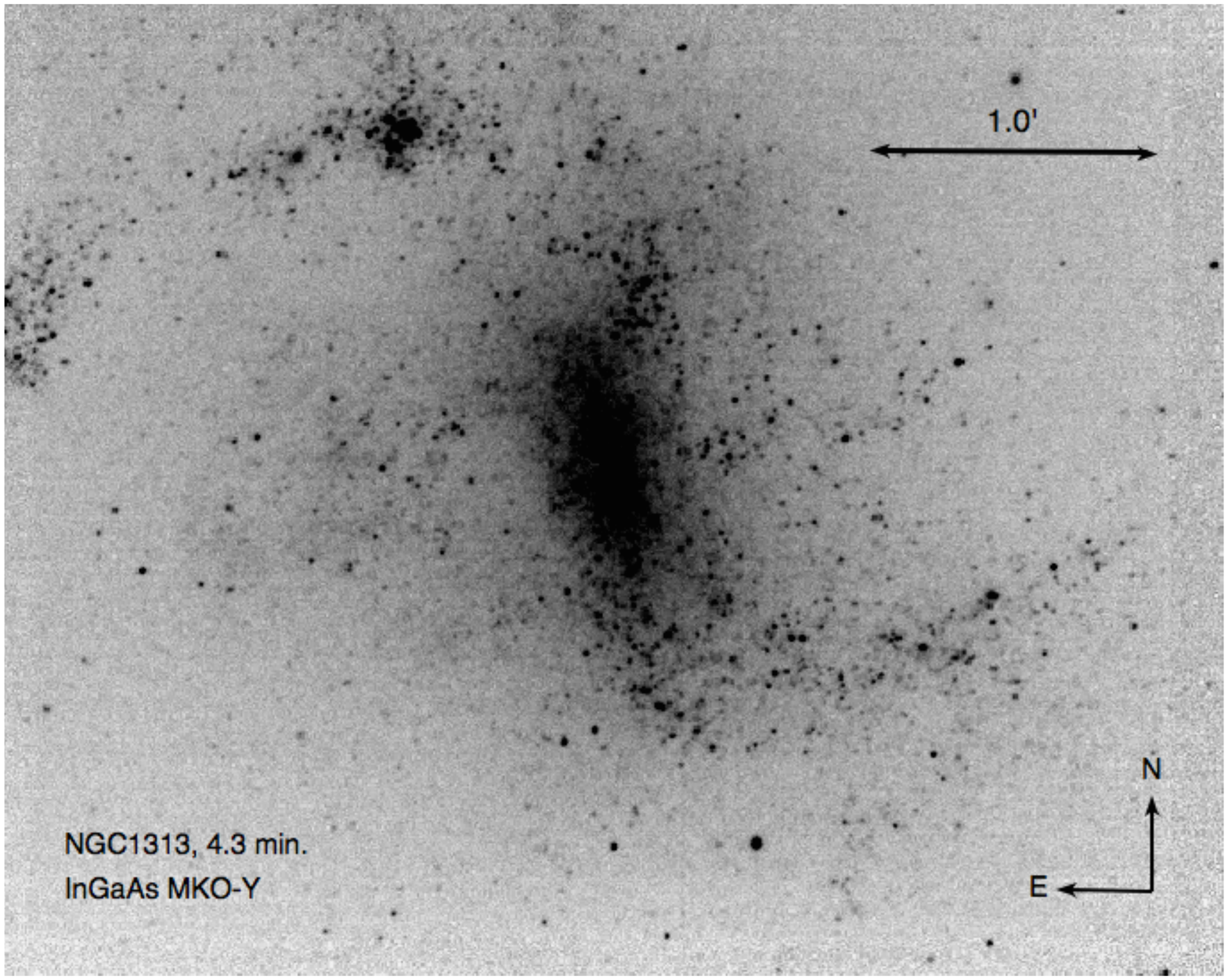}{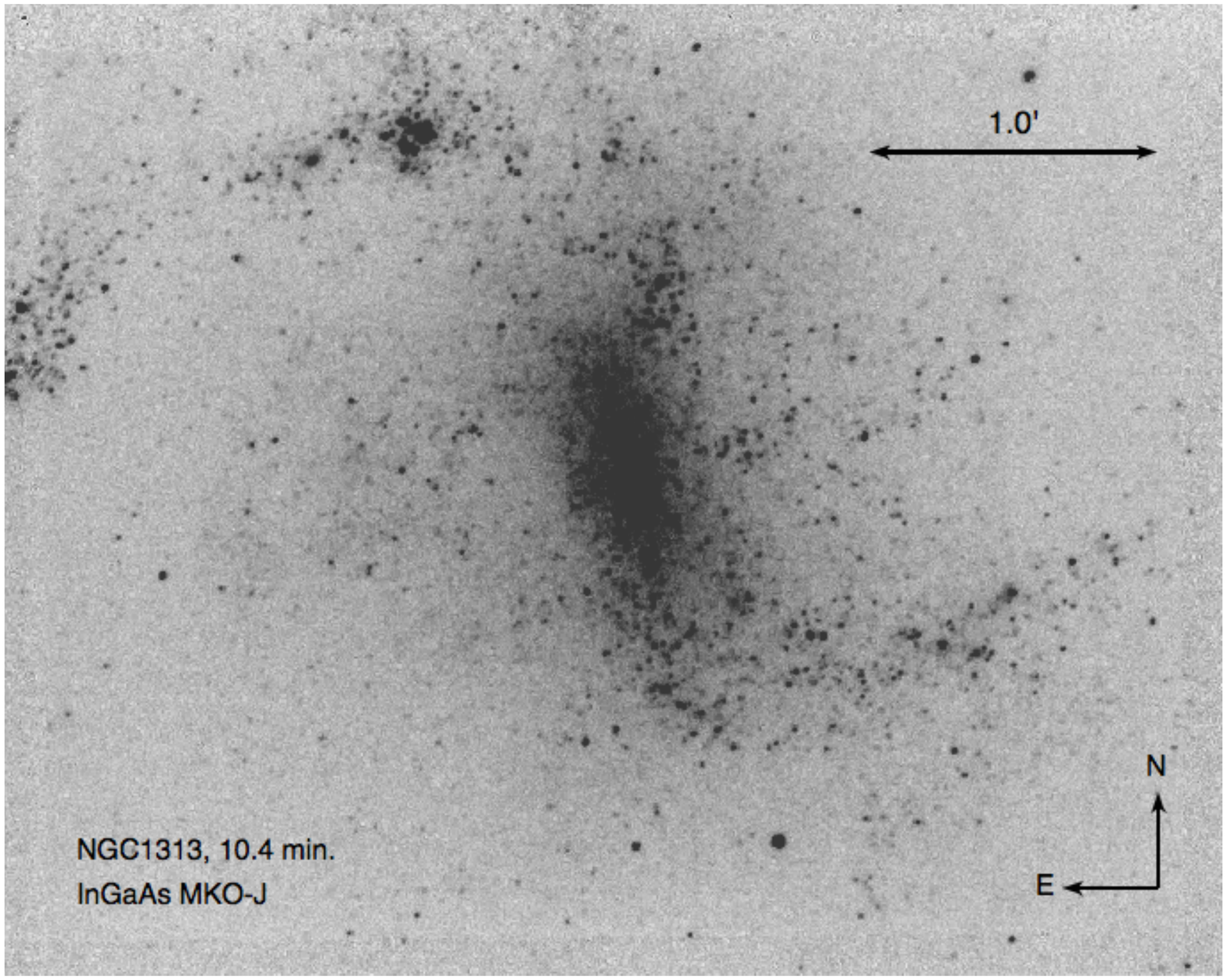}}
  \caption{\label{fig:ngc1300}\small Y and J images of nearby galaxy
    NGC1300 (top row) and the SPIRITS survey target NGC1313 (bottom
    row). }
\end{figure*} 

A more sophisticated estimate of sensitivity or completeness could be
made by injecting simulated objects into the data and testing the
efficacy of Sextractor at recovering these objects. However the
present analysis provides a sufficiently general estimate of
photometric speed to evaluate potential survey science programs,
described below.

\subsection{Representative Science Observations} \label{sec_backgrounds}

We observed several selected science targets chosen to represent a
range of possible observational programs that would benefit from
background-limited IR photometry in the $Y$ through $H$ bands at
reduced cost and complexity. These specifically include (a)
observations of nearby galaxies typical of astrophysical transient
searches, (b) deep images of fields at cosmological distances to
search for distant quasars and clusters, and (c) observations of
transiting exoplanets in the near-infrared.

\subsubsection{Nearby Galaxies, including IR Transient Survey Objects}

We observed two galaxies from the local universe: NGC1300 and NGC1313
(Figure \ref{fig:ngc1300}).  The latter object in particular was
targeted because it is used for a synoptic IR survey of obscured
transients using the Spitzer Space Telescope.  NGC1300 was observed
for 7.6 minutes in $Y$ and $17.2$ minutes in $J$, whereas NGC1313 was
observed for 4.3 and 10.4 minutes in the same filters, respectively.
The unusual exposure times reflect rounding to the nearest clock time
for an even number of ramp frames, and were varied throughout the run
as we refined our observing strategy.

\begin{figure*}[!t]
  \begin{center}
    \begin{tabular}{c} 
      \includegraphics[height=5cm]{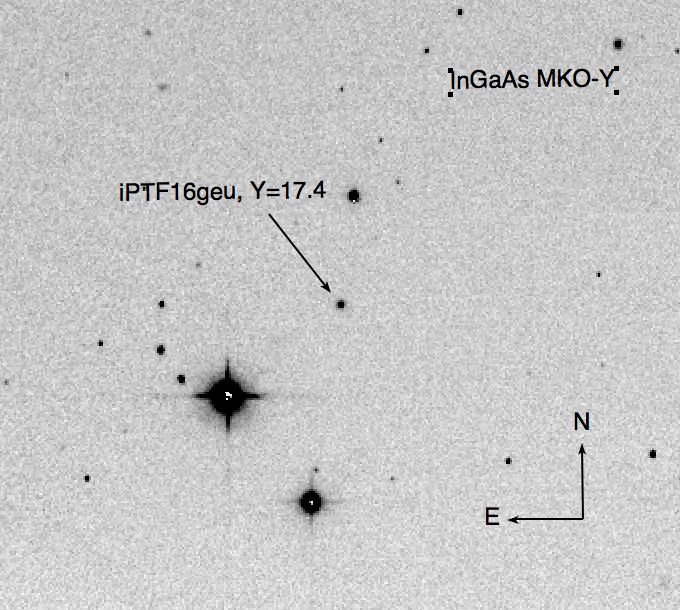}
      \includegraphics[height=5cm]{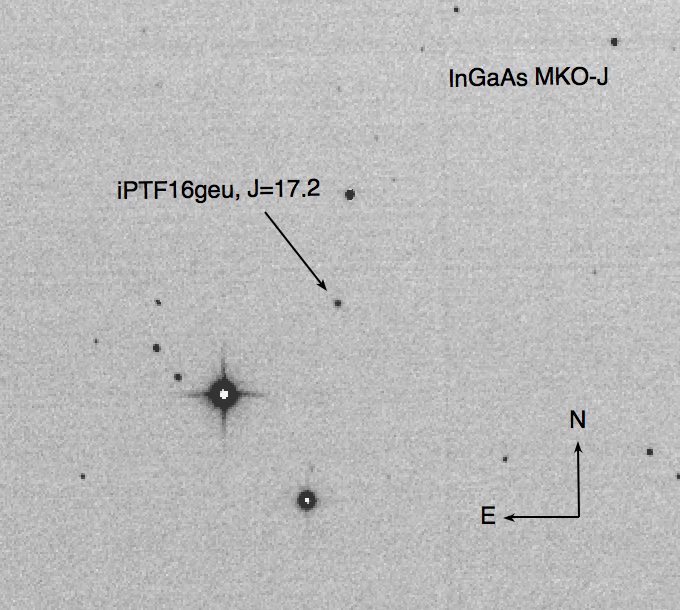}
      \includegraphics[height=5cm]{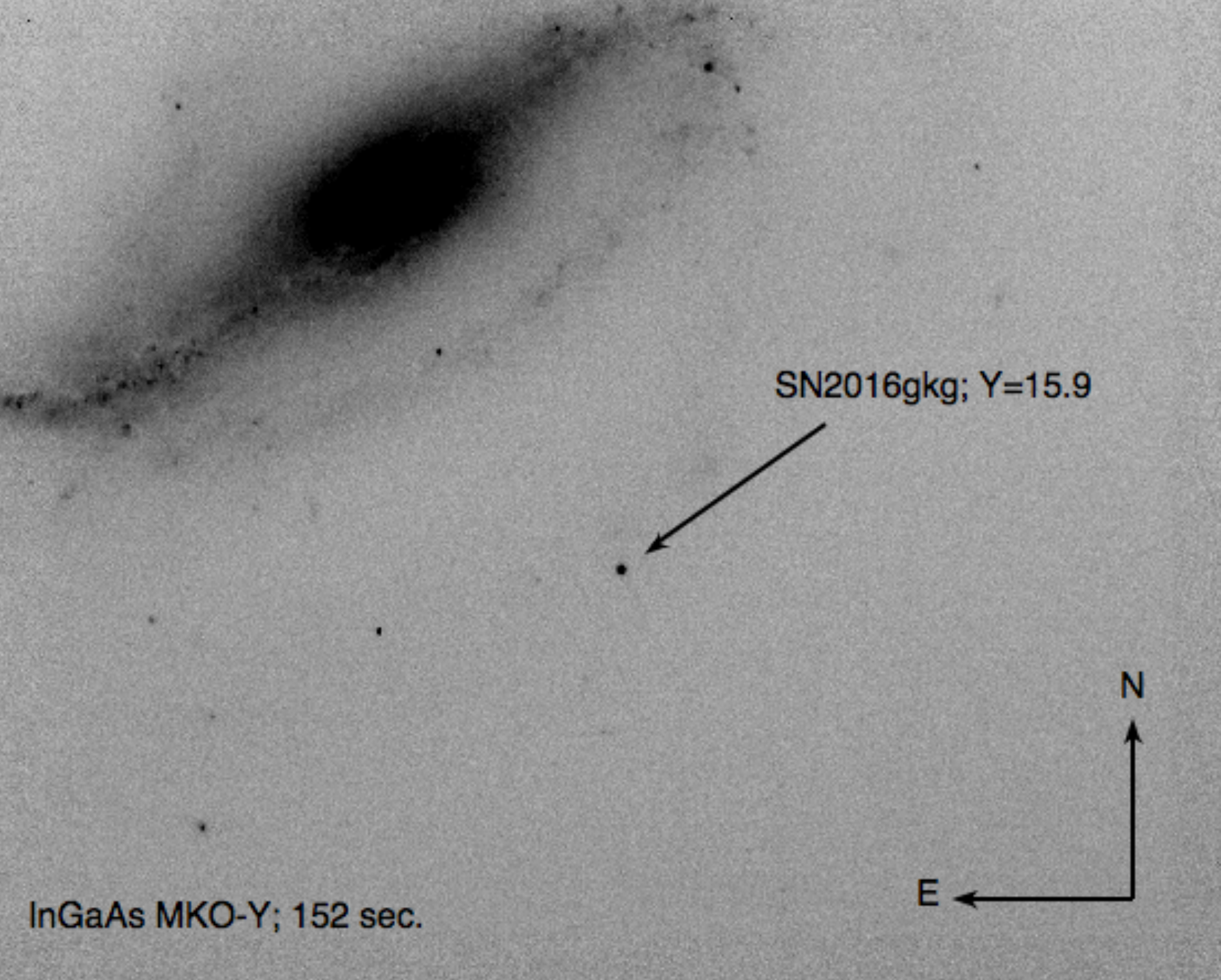}\\
    \end{tabular}
  \end{center}
  \caption{\label{fig:iPTF16geu}$Y$ (left, 391 seconds)
    and $J$ (center, 326 seconds) band images of the gravitationally
    lensed Type Ia supernova iPTF16geu. At right is a Y band image of
    SN2016gkg (152 seconds).}
\end{figure*} 

\subsubsection{Supernovae and Explosive Transients}

A primary motivation for building a wide-field IR camera is to pursue
time-domain science in the $Y$ through $H$ bands.  Accordingly, we
observed two supernovae that were visible during the run as a
demonstration.

Our first target was iPTF16geu, a gravitationally lensed Type Ia
supernova \citep{goobar17}.  Figure \ref{fig:iPTF16geu} shows $Y$ and
$J$ band images of the field, with 391 and 326 second integrations,
respectively, taken on 2016 November 13 (UT00:22:27 for $Y$ and
UT00:53:58 for $J$).  We measure a total apparent magnitude of
$Y_{AB}=17.4$ at $55\sigma$ significance, and $J_{AB}=17.2$ at
$38\sigma$.  Although the data were taken in good seeing conditions,
we were not able to resolve individual components of the lens.

Our other supernova target is the Type IIb source SN2016gkg
\citep{2016gkg_nature}, observed 152 seconds in $J$ and 206 seconds in
$Y$ (Figure \ref{fig:iPTF16geu}).  We detect the supernova with
$Y_{AB}=15.9$ at $214\sigma$ and $J_{AB}=16.12$ at $205\sigma$.

\subsubsection{High-Redshift QSOs}

A key static-sky application for near-IR imagers is the search for
high redshift QSOs in deep data sets.  To test whether the InGaAs
sensor is sensitive to currently known high-$z$ populations, we
observed the known $z=6.31$ quasar ATLAS J025.6821-33.4627
\citep{carnall_atlas}.  This object was (somewhat unusually) selected
from $z-W1$ (WISE) colors, so its near-IR magnitudes were not reported
in the literature.

\begin{figure}[b]
  \plotone{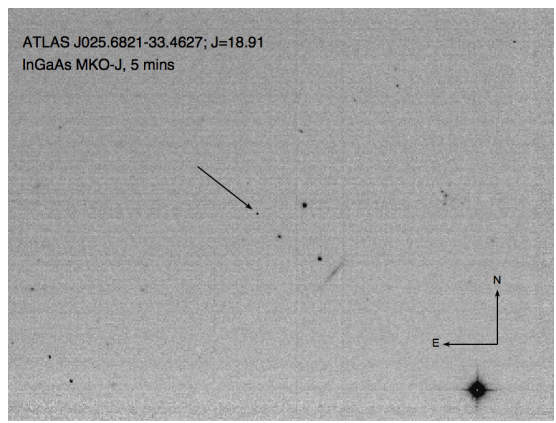}
  \caption{\label{fig:j025}Image of the $z=6.31$ quasar
    ATLAS J025.6821-33.4627.  This source is clearly detected at
    $20\sigma$ signficance for $J_{AB}=18.91$ in a $\sim 5$ minute
    integration.}
\end{figure}

Figure \ref{fig:j025} shows a cutout of the $J$ band image, which was
observed for 326 seconds. We detect a source at the expected location
of the QSO with $J_{AB} = 18.91$ and $20\sigma$ significance. The
quasar was also observed for 717 seconds in $Y$ and a source is
clearly visible in the data.  However poor image registration (caused
by a low star count in the high-latitude field) complicates photometry
and requires further refinement to produce a well-calibrated
measurement.  

\subsubsection{Faint/High-Redshift Galaxies and/or Clusters}

%

\begin{figure*}[t]
  \plottwo{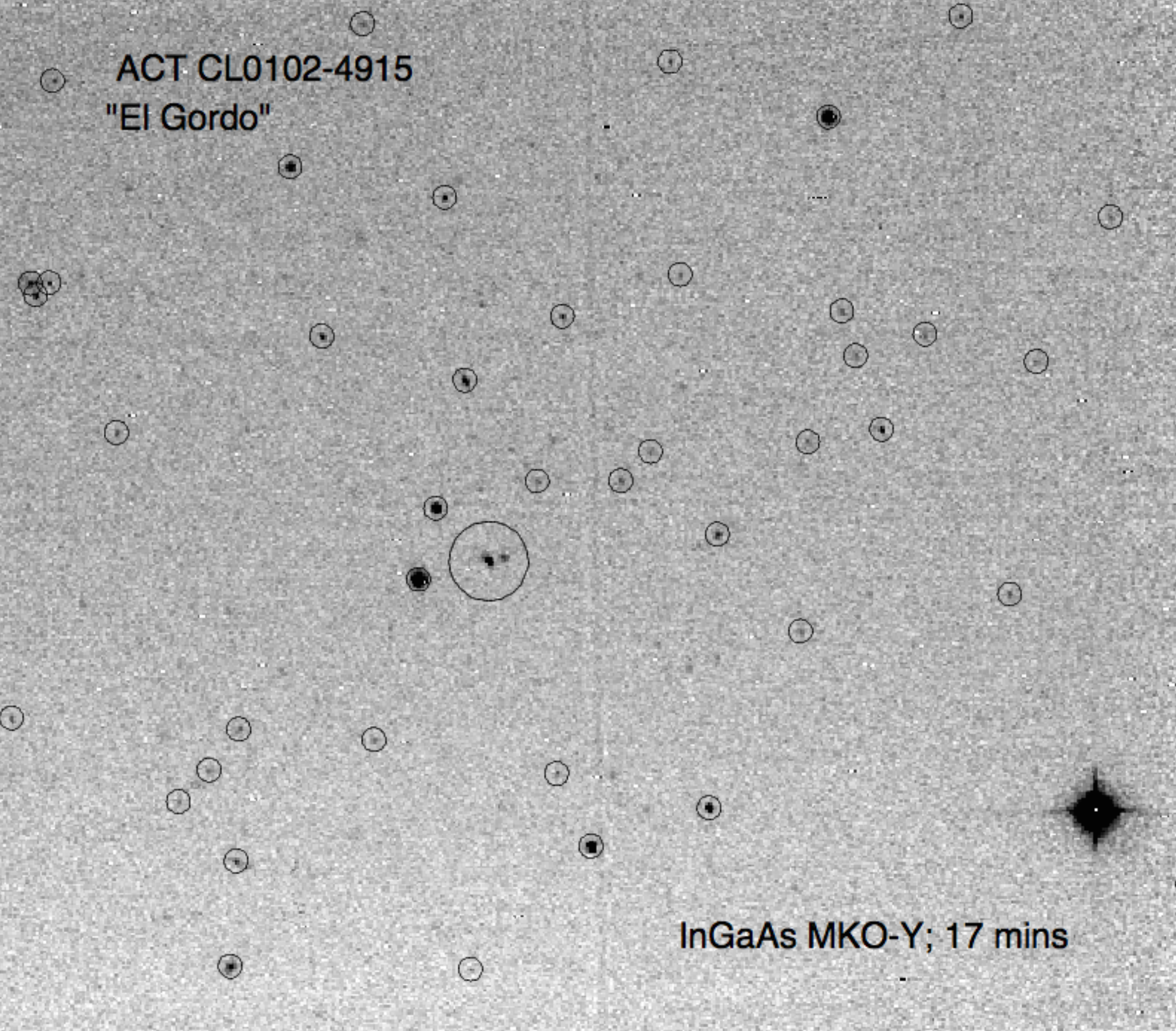}{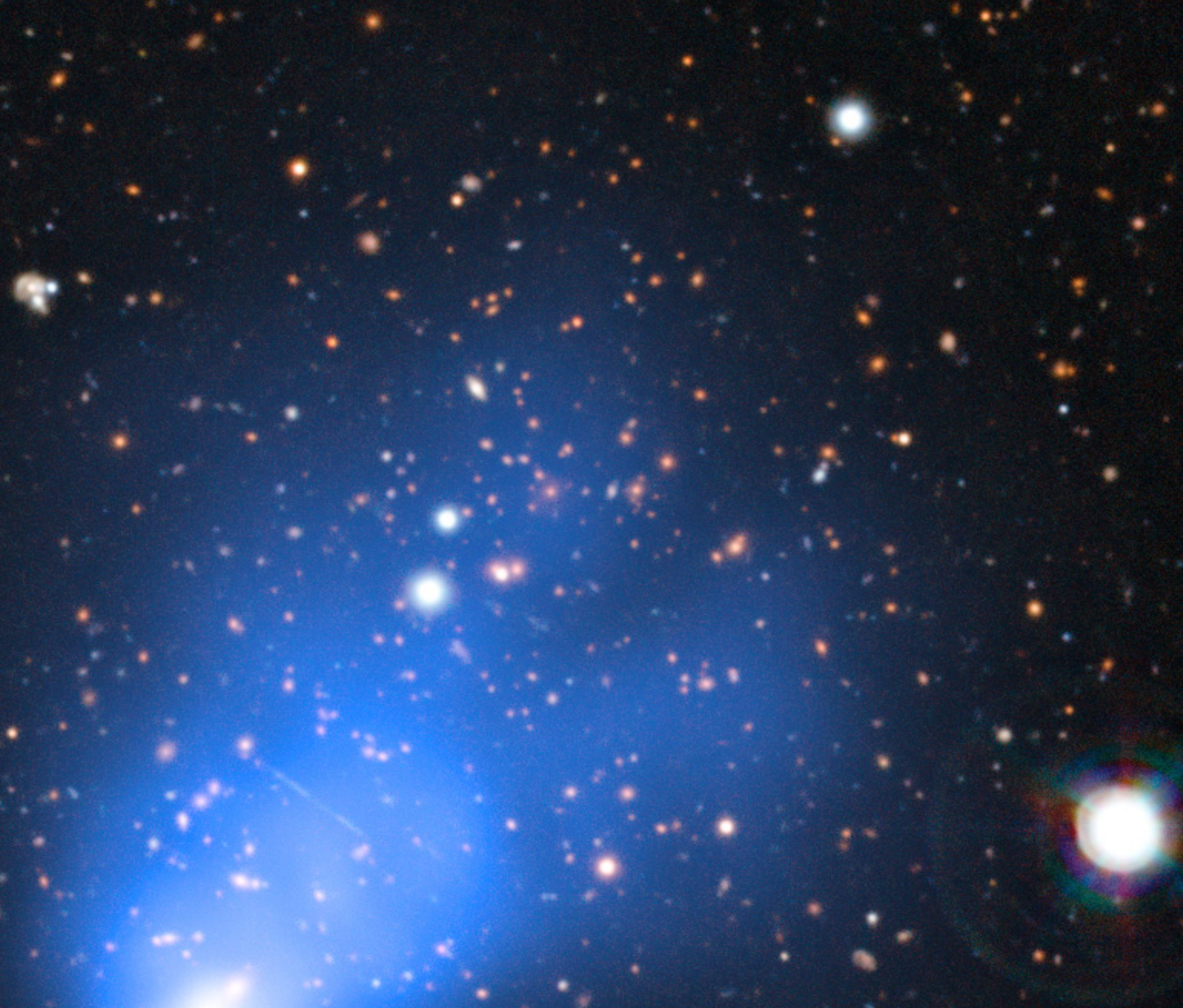}
  \caption{\label{fig:elgordo}Left: 17 minute $Y$ band image of the
    galaxy cluster ACT CLJ0102-4915 \citep[``El
      Gordo'']{menanteau2012} at $z=0.87$. The Brightest Cluster
    Galaxy is indicated at center, and has $Y\approx 18$. The majority
    of the other $5\sigma$ detections indicated are at $Y\approx
    21.0-22.5$.  Right: Optical image of the same field with VLT/SOAR
    for reference, with Chandra map overlaid in blue
    \citep{menanteau2012}.}
\end{figure*}

To push photometric depth, we constructed a 17-minute $Y$-band stack
of the field containing the $z=0.87$ galaxy cluster ACT-CL J0102-4915
\citep[nicknamed ``El Gordo'';][]{menanteau2012}. Using the zero points
listed in Table 1, we extracted magnitudes for all sources detected
with $\ge 5\sigma$ signifiance---these are circled in Figure
\ref{fig:elgordo}. Although the extremely luminous Brughtest Cluster
Galaxy (BCG) is well above the noise floor at $Y\approx 18$, we also
detect numerous objects from the red sequence at $y=21.5-22.5$, our
approximate detection limit.

This observation, together with the $z=6.3$ QSO observations presented
in 5.3.3 suggests that wide-field InGaAs mosaics can deliver
sufficient imaging depth to survey, discover and charaterize objects
at cosmological distances.

\subsubsection{Exoplanet Transit}

Our early studies with InGaAs were partially motivated by an interest
in using the sensors for exoplanetary transit surveys around low-mass
stars, including L and T dwarfs which are bright in the $J$ band.  Our
earlier work \citep{sullivan_ap640c,sullivan_ap1121} included
laboratory tests of photometric stability, but did not present on-sky
detections of transit events.

We did not schedule the DuPont run explicitly for optimal observation
of an exoplanet transit, but a database
search\footnote{http://var2.astro.cz/ETD/} revealed a small number of
partial transits which were visible from Las Campanas during our run.
Because of the InGaAs camera's small field size, we focused our search
on targets with bright nearby comparison stars.

We observed the newly discovered Hot Jupiter HATS34b, a 0.94
Jupiter-mass planet orbiting a $V=13.9/J=12.5$ (Vega) host star of
$T_{eff}=5380$K.  The object orbits with $P=2.1$ days and was visible
through ingress and transit during last portion of the night.  Its
field includes a second comparison star 2.5 magnitudes brighter than
HATS34, at a projected distance of 163\arcsec.  Sunrise prevented
observations of the transit egress and re-establishment of a
post-transit photometric baseline.  The transit depth reported in the
discovery paper \citep{deValBorro} is 13.4 mmag, or 1.2\%.

So as not to saturate either star, we obtained short individual
exposures of $t=1.03$ seconds, with the telescope slightly defocused.
The telescope was guided throughout the sequence by an off-axis probe
provided by the observatory, to maintain stable positioning of objects
on the array.

We extracted fluxes of the science target and reference star using
Sextractor in strict aperture photometry mode, with an aperture
diameter of 15 pixels (6\arcsec).  No attempts were made to optimize
the photometric extraction parameters or aperture.

\begin{figure} [t]
  \plotone{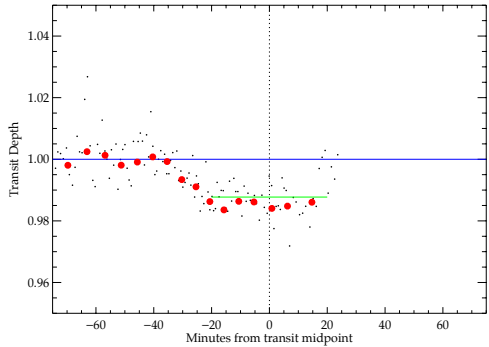}
  \caption{\label{fig:transit}$J$ band light curve of
    HATS-34b recorded over the partial transit of UT2016 November 12.
    Black dots indicate 11-exposure bins, for effective exposure time
    of 11.33 seconds.  The solid red cirles are more heavily binned
    into intervals of 82.4 seconds.}
\end{figure}

Figure \ref{fig:transit} shows the resulting lightcurve, constructed
from differential photometry between the target and reference stars.
To reduce shot noise, we bin the counts from multiple exposures by
adding the photometric fluxes with a top hat window of varying width,
to verify the scaling of noise reduction.  We center around the known
time of transit calculated from the HATS34b discovery paper.  The
black solid points are averages of 11 exposures, or 11.33 seconds,
whereas the red points average 80 exposures for 82.4 seconds.

Even in the 11 second averages, there is a clear transit detection
with a depth consistent with the value reported by the HATS team (the
expected optical transit depth is indicated with a light green line).
In the 82-second bins the detection is highly significant, with a
transit depth slightly larger than the 1.2\% predicted by HATS, but
within the margin of error.

Our main objective was not to fit a new transit lightcurve and
re-derive the orbital properties of HATS34b.  We simply demonstrate
that InGaAs sensors are capable not just of deep photometry and
detection of explosive transients, but also of precision photometry at
the milli-magnitude level, over long time baselines.  These data were
obtained in the $J$ band where the sky is brighter and more variable
in emission and transparency than the optical.  It suggests that with
proper attention to noise, stability, and observation InGaAs cameras
can offer an affordable alternative to costly HgCdTe arrays for IR
transit observation---particularly if a large format is not required.

\section{Discussion}

\subsection{Suitability for Synoptic Infrared Kilonova Surveys or LSST Synergy}

Wide-field near-IR imagers are potentially attractive survey
instruments to search for and characterize the electromagnetic (EM)
counterparts of binary neutron star (BNS) mergers detected via
gravitational waves.  Pioneering work by \citet{lipaczynski} and
\citet{metzger} developed predictions for an isotropic EM ``kilonova''
signature powered by radioactive decay of heavy elements synthesized
via neutron capture.  Using opacity tables for heavy elements in the
Lanthanide series, \citet{kasen} and \citet{barnes} further predicted
that the most long-lived emission from these events will emerge in the
$Y$ through $H$ bands. This results from an increased opacity at
optical wavelengths relative to the iron peak elements seen in
conventional supernova photospheres.

The recent discovery of an apparent kilonova \citep[or similar
  ``macronova'';][]{metzger2015,kasen2015} associated with GW170817
\citep{ligo_ns} supports this basic picture, although the observed optical
counterpart is far brighter than expected for material whose opacity
is dominated by the heaviest $r$-process elements.  This event
therefore offers an ideal opportunity to assess the relative merits of
wide-field optical versus IR imagers for followup of anticipated
future BNS events.

\begin{figure*}[t]
  \plotone{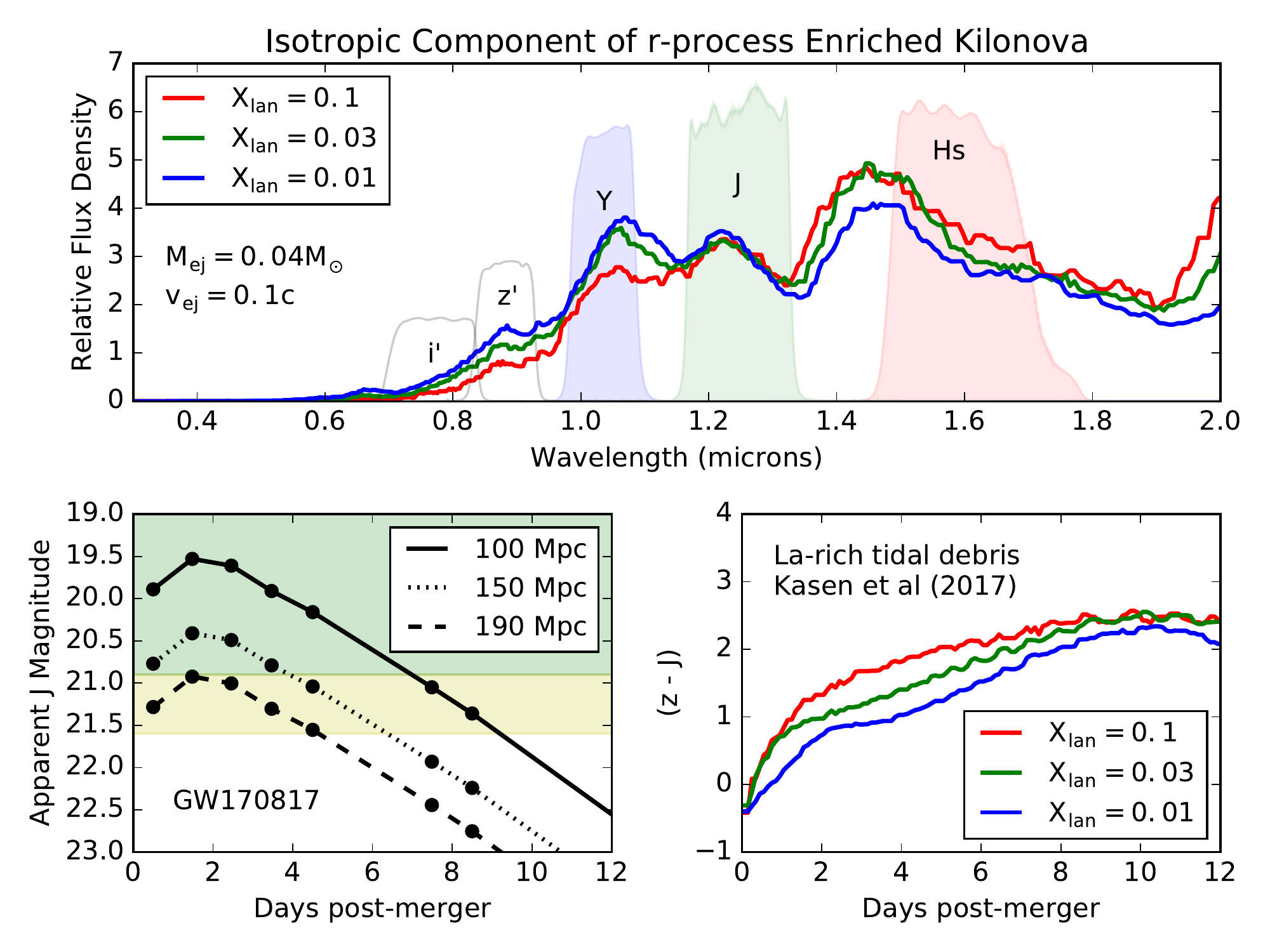} 
  \caption{\label{fig:kilonova}\small{\it Top:} Spectral
    models of kilonova emission at $t=2$ days
    post-merger \citep{kasen_nature}. All models have an ejected mass of
    $0.04M_\odot$ at velocity $0.1c$, similar to the favored
    parameters of GW170817 but with varying lanthanide fractions from
    1-10\%. All models have been normalized to a common flux in $J$,
    and filter curves show the bandpass of MKO filters weighted by
    InGaAs quantum efficiency.  {\it Bottom left:} Observed $J$-band
    light curve of GW170817 projected to distances between 100-190
    Mpc---the expected horizon for Advanced
    LIGO\citep{ligo_o4}. The green shaded region corresponds to a
    5-minute integration depth, whereas the yellow region corresponds
    to 10 minutes integration. {\it Bottom Right:} Temporal evolution
    of $(z-J)$ color in the models from the top panel as the EM
    counterpart fades in the first 2 weeks after merger. Rapid fading
    together with strong reddning is a characteristic signature of
    lanthanide-rich merger remnants.}
\end{figure*}

Figure \ref{fig:kilonova} shows the emergent spectrum predicted by the
models of \citet{kasen_nature} for varying mass fractions of
lanthanide elements in the post-merger ejecta, all at $t=2$
days. Kasen's full model of GW170817 required two components: one
polar outflow containing light $r$-process elements moving at high
speed, and one isotropic component of emission from $0.04M_\odot$ of
heavy $r$-process material moving at $v_{ej}=0.1c$. The latter,
isotropic component is shown in the top panel of the figure. Filter
curves for MKO $Y, J$, and $H_s$ (conventional $H$ multiplied by the
InGaAs QE cutoff) illustrate that the flux density of lanthanide-rich
tidal debris is higher at near-IR wavelengths than red optical bands,
partially offsetting the built-in cost and heritage advantage of
existing CCD-based search strategies.

On the other hand, the fast outflow has been associated with a
short-lived ($t\sim 1$ day), blue EM transient, whose exact mechanism
is not fully constrained nor is it known whether such short-lived,
bright optical counterparts are generic to kilonova events or are
orientation-dependent. Two possible early models include a gamma-ray
burst seen at an off-axis angle \citep{offax_jet}, and shock-heated
gas from a cocoon of material surrounding a relativistic jet that has
either been choked off before emerging, or has just emerged but is
seen off-axis \citep{nakar2012,piro,kasliwal_nsns}. Late-time radio
observations appear to favor such cocoon models \citep{mooley}.

The IR emission driven by radioactive decay of heavy lanthanides
should be largely isotropic and visible for a week or more
\citep{metzger,kasen}, it is a ubiquitous prediction of the model and
should be visible for any viewing orientation.  The bottom left panel
of Figure \ref{fig:kilonova} plots the observed $J$ band light curve
of GW170817 \citep{drout,cowperthwaite_ns,villar}, now projected to
distances of 100, 150 and 190 Mpc. The latter corresponds to the
expected horizon distance for NS-NS merger detections in LIGO's fourth
observing run \citep{ligo_o4}.

Our demonstration camera would detect the fading counterpart of
GW170817 in $J$ using 5-minute survey integrations (green shaded
region) at $5\sigma$ significance for a week after merger at 100 Mpc,
or 4 days at 150 Mpc, and would just miss a detection in 5 minutes at
LIGO's maximum distance.  If 10-minute mapping integrations are used
(yellow shaded region), the transient is visible for 9 days at 100
Mpc, and 5 days at 190 Mpc.  The $(z-J)$ color of the slow, heavy wind
rises steeply with time, exhibiting 1-2 magnitudes of reddning over
the first few days following merger (Figure 13, bottom right).

With these plots and the sensitivity curves from Figure
\ref{fig:depth}, we can assess the relative merits of an optical
survey in the $i^\prime$ or $z^\prime$ bands versus an infrared InGaAs
camera in $Y$, $J$ or $H$.

For concreteness, we assume a fiducial 1 square degree FoV tiled with
InGaAs, consistent with an under-filled focal plane on the 2.5-meter
DuPont or SDSS telescopes (which have a $\sim 3.1$ square degree
field), although similar scaling arguments can be developed for
smaller apertures. If we further assume a 10-minute integration
cadence, Figure \ref{fig:depth} indicates a $5\sigma$ depth of
$J_{AB}=21.5-21.7$.  At this speed, one could map $\sim 6$ square
degrees per hour with sufficient sensitivity to detect the EM
counterpart of GW170817 at 150 Mpc for 7 days.  At the actual distance
of 40 Mpc for GW170817, the $r$-process peak would be visible for over
two weeks in 10 minute exposures.  At the edge of the Advanced LIGO
design horizon of 200 Mpc, it would be visible for 5 days.

Simulations suggest that in the era of two GW detectors (i.e. Advanced
LIGO O3), roughly half of all BNS mergers will be localized within
$150$ square degrees \citep{chen_holz_15}, and only 10\% will have
localizations of 50 square degrees or better. However if the
positional region of uncertainty is above the horizon for 3-4 hours
per night, one may survey 18-24 square degrees per day to
$J_{AB}=21.5$. This is sufficient to tile the most well-localized 10\%
of events in 2 nights, or survey even unfavorable two-detector error
contours in four to five nights (during which time the IR transient is
still visible throughout the full 190 Mpc volume).

The same simulations find that inclusion of VIRGO \citep{virgo}
detections during O4 will localize $\sim 70$\% of events within
approximately 5 square degrees.  This region could be surveyed with
1-2 hour cadence to the same depth, but would more likely be integrated
to $\ge$1-hour exposure depth reaching $J_{AB}=22.5$ or fainter.  This
would allow IR instruments to identify mergers with lower ejected mass
yields or velocities, consistent with prior expectations for BNS
mergers \citep{kasen2015}.  It also provides more margin for events
that are dust-obscured or inconveniently placed on the sky. The lower
cost of InGaAs could enable blind searches for NS-NS merger
counterparts directly in the IR, in parallel with similar searches at
optical wavelengths.

Because the EM counterpart of GW170817 was discovered in the optical
and existing CCD imagers of larger entendue will already be searching
for the same counterparts, one must consider whether there is added
value in contemporaneous IR searches using smaller apertures and/or
fields. Indeed a rapidly fading optical transient was the first and
brightest signal seen from this event on the ground.

There are three possible motivations for pairing deep optical searches
with wide-field IR mapping (as opposed to post-discovery followup).
First, the radioactively heated IR transient is widely believed to be
isotropic, while the angular dependence of the UV-optical radiation is
unconstrained at present. An accounting of the fractional contribution
of UV-optical transients to the parent population of IR-trggered
events will define the geometry and uniformity of the EM mechanisms.

Second, a generic signature of heated-cocoon models is a rapid optical
transient that fades by over 1 magnitude per day, concomitant with a
rise in the $J$ and $H$ bands over similar time scales---in other
words, all bands bluer than $J$ fade
continuously. \citet{cowperthwaite_survey} demonstrated that optical
searches can map regions of $\sim 50$ square degrees to
$i^\prime=22.5$, sufficient to detect a rapid UV-optical transient
similar to GW170817 for 3-4 days. Optical searches would not detect
emission from the isotropic $r$-process material; they require
favorably oriented jets, cocoons, or disk winds to successfully
identify EM counterparts.

However \citet{cowperthwaite_survey} also find that the optical
searches exhibit a false-positive rate of $\sim 2$ unrelated
transients per square degree in their blind search; after applying
priors on color this rate is reduced to $\sim 1$ degree$^{-2}$ if
kilonovae are assumed to be intrisically blue, or $\sim 0.15$
deg$^{-2}$ if they are intrisically red as in $r$-process events. A
blind search of $\sim 50$ square degrees would therefore yield between
7 and 100 contaminants depending on priors applied. Early-time
monitoring in the $J$ or $H$ bands could establish the simultaneous IR
brightening and sustained luminosity unique to NS-NS binary mergers,
separating BNS events from unrelated foregrounds.

Third, the long duration of EM signals in the infrared offers an
extended window over which to monitor events, in contrast to
high-etendue optical campaigns which cannot always encumber the
resources of large telescopes for followup on timescales of weeks.

It is projected that aLIGO and VIRGO may discover NS binaries at a
rate of 1/month at design sensitivity, such that followup of these
events consumes a substantial portion, but not all of small
observtories' time allocations. During times when targeted followups
are not underway, such a survey instrument could perform a dedicated
wide-field time-domain survey in the IR, a program which has not been
undertaken to date largely on account of sensor costs.

\section{Conclusions}\label{sec_conclusions}

We report a series of tests from a prototype InGaAs camera deployed on
the 2.5-meter DuPont telescope. On an aperture of this size we find
that the AP1121 can deliver sky-photon limited noise performance in
the $J$ band with $0.4\arcsec$ pixels, and has roughly equal
contributions from sky and dark current in $Y$, in operating
conditions at $T=-40$C with no cold stop.  A modest engineering effort
was needed to reduce read noise through non-destructive sampling (not
generally available on commercial cameras), to levels where it is
quickly exceeded by shot noise from the sky. This indicates that for
broadband imaging applications not requiring the $K$ band, InGaAs can
be competetive with HgCdTe at substantially reduced cost.

On a 2.5-meter telescope, we measure photometric zero points of
$24.5-25.3$ magnitudes in the $Y$ and $J$ bands, demosntrating
sufficient sensitivity to image transient sources in the local
universe, the red sequence of a $z=0.87$ galaxy cluster, and a $z=6.3$
QSO, typical of static-sky survey targets.

While these devices have less astronomy heritage, their cost savings
could make them attractive for wide-field survey instruments on medium
sized apertures, or alternatively as low-cost IR photometers on
1-meter apertures with pixels of $\sim 1$\arcsec ~or larger.

\acknowledgements 

We gratefully acknowledge support from the Kavli Research Investment
Fund at MIT for early development of custom InGaAs detector cameras
and their associated hardware and optics. We also thank the technical
staff of Carnegie Observatories and Las Campanas for their logistical
support of a complex shipment to Chile and successful installation and
operation on the telescope, as well as to the scientific staff and
Director for arranging time and support for our observations on the
DuPont telescope. 

\facility{DuPont}

\bibliography{ingaas} 

\begin{thebibliography}{40}
\expandafter\ifx\csname natexlab\endcsname\relax\def\natexlab#1{#1}\fi

\bibitem[{{Abbott} {et~al.}(2016){Abbott}, {Abbott}, {Abbott}, {Abernathy},
  {Acernese}, {Ackley}, {Adams}, {Adams}, {Addesso}, {Adhikari}, \&
  et~al.}]{ligo_o4}
{Abbott}, B.~P., {et~al.} 2016, Living Reviews in Relativity, 19, 1

\bibitem[{Abbott {et~al.}(2017)Abbott, Abbott, Abbott, Acernese, Ackley, Adams,
  Adams, Addesso, Adhikari, Adya, Affeldt, Afrough, Agarwal, Agathos, Agatsuma,
  Aggarwal, Aguiar, Aiello, Ain, Ajith, Allen, Allen, Allocca, Altin, Amato,
  Ananyeva, Anderson, Anderson, Angelova, Antier, Appert, Arai, Araya, Areeda,
  Arnaud, Arun, Ascenzi, Ashton, Ast, Aston, Astone, Atallah, Aufmuth, Aulbert,
  AultONeal, Austin, Avila-Alvarez, Babak, Bacon, Bader, Bae, Bailes, Baker,
  Baldaccini, Ballardin, Ballmer, Banagiri, Barayoga, Barclay, Barish, Barker,
  Barkett, Barone, Barr, Barsotti, Barsuglia, Barta, Barthelmy, Bartlett,
  Bartos, Bassiri, Basti, Batch, Bawaj, Bayley, Bazzan, B\'ecsy, Beer, Bejger,
  Belahcene, Bell, Berger, Bergmann, Bernuzzi, Bero, Berry, Bersanetti,
  Bertolini, Betzwieser, Bhagwat, Bhandare, Bilenko, Billingsley, Billman,
  Birch, Birney, Birnholtz, Biscans, Biscoveanu, Bisht, Bitossi, Biwer,
  Bizouard, Blackburn, Blackman, Blair, Blair, Blair, Bloemen, Bock, Bode,
  Boer, Bogaert, Bohe, Bondu, Bonilla, Bonnand, Boom, Bork, Boschi, Bose,
  Bossie, Bouffanais, Bozzi, Bradaschia, Brady, Branchesi, Brau, Briant,
  Brillet, Brinkmann, Brisson, Brockill, Broida, Brooks, Brown, Brown, Brunett,
  Buchanan, Buikema, Bulik, Bulten, Buonanno, Buskulic, Buy, Byer, Cabero,
  Cadonati, Cagnoli, Cahillane, Calder\'on~Bustillo, Callister, Calloni, Camp,
  Canepa, Canizares, Cannon, Cao, Cao, Capano, Capocasa, Carbognani, Caride,
  Carney, Carullo, Casanueva~Diaz, Casentini, Caudill, Cavagli\`a, Cavalier,
  Cavalieri, Cella, Cepeda, Cerd\'a-Dur\'an, Cerretani, Cesarini, Chamberlin,
  Chan, Chao, Charlton, Chase, Chassande-Mottin, Chatterjee, Chatziioannou,
  Cheeseboro, Chen, Chen, Chen, Cheng, Chia, Chincarini, Chiummo, Chmiel, Cho,
  Cho, Chow, Christensen, Chu, Chua, Chua, Chung, Chung, Ciani, Ciolfi,
  Cirelli, Cirone, Clara, Clark, Clearwater, Cleva, Cocchieri, Coccia, Cohadon,
  Cohen, Colla, Collette, Cominsky, Constancio, Conti, Cooper, Corban, Corbitt,
  Cordero-Carri\'on, Corley, Cornish, Corsi, Cortese, Costa, Coughlin,
  Coughlin, Coulon, Countryman, Couvares, Covas, Cowan, Coward, Cowart, Coyne,
  Coyne, Creighton, Creighton, Cripe, Crowder, Cullen, Cumming, Cunningham,
  Cuoco, Dal~Canton, D\'alya, Danilishin, D'Antonio, Danzmann, Dasgupta,
  Da~Silva~Costa, Dattilo, Dave, Davier, Davis, Daw, Day, De, DeBra, Degallaix,
  De~Laurentis, Del\'eglise, Del~Pozzo, Demos, Denker, Dent, De~Pietri,
  Dergachev, De~Rosa, DeRosa, De~Rossi, DeSalvo, de~Varona, Devenson,
  Dhurandhar, D\'{\i}az, Dietrich, Di~Fiore, Di~Giovanni, Di~Girolamo,
  Di~Lieto, Di~Pace, Di~Palma, Di~Renzo, Doctor, Dolique, Donovan, Dooley,
  Doravari, Dorrington, Douglas, Dovale~\'Alvarez, Downes, Drago,
  Dreissigacker, Driggers, Du, Ducrot, Dudi, Dupej, Dwyer, Edo, Edwards,
  Effler, Eggenstein, Ehrens, Eichholz, Eikenberry, Eisenstein, Essick,
  Estevez, Etienne, Etzel, Evans, Evans, Factourovich, Fafone, Fair, Fairhurst,
  Fan, Farinon, Farr, Farr, Fauchon-Jones, Favata, Fays, Fee, Fehrmann, Feicht,
  Fejer, Fernandez-Galiana, Ferrante, Ferreira, Ferrini, Fidecaro, Finstad,
  Fiori, Fiorucci, Fishbach, Fisher, Fitz-Axen, Flaminio, Fletcher, Fong, Font,
  Forsyth, Forsyth, Fournier, Frasca, Frasconi, Frei, Freise, Frey, Frey,
  Fries, Fritschel, Frolov, Fulda, Fyffe, Gabbard, Gadre, Gaebel, Gair,
  Gammaitoni, Ganija, Gaonkar, Garcia-Quiros, Garufi, Gateley, Gaudio, Gaur,
  Gayathri, Gehrels, Gemme, Genin, Gennai, George, George, Gergely, Germain,
  Ghonge, Ghosh, Ghosh, Ghosh, Giaime, Giardina, Giazotto, Gill, Glover, Goetz,
  Goetz, Gomes, Goncharov, Gonz\'alez, Gonzalez~Castro, Gopakumar, Gorodetsky,
  Gossan, Gosselin, Gouaty, Grado, Graef, Granata, Grant, Gras, Gray, Greco,
  Green, Gretarsson, Groot, Grote, Grunewald, Gruning, Guidi, Guo, Gupta,
  Gupta, Gushwa, Gustafson, Gustafson, Halim, Hall, Hall, Hamilton, Hammond,
  Haney, Hanke, Hanks, Hanna, Hannam, Hannuksela, Hanson, Hardwick, Harms,
  Harry, Harry, Hart, Haster, Haughian, Healy, Heidmann, Heintze, Heitmann,
  Hello, Hemming, Hendry, Heng, Hennig, Heptonstall, Heurs, Hild, Hinderer, Ho,
  Hoak, Hofman, Holt, Holz, Hopkins, Horst, Hough, Houston, Howell, Hreibi, Hu,
  Huerta, Huet, Hughey, Husa, Huttner, Huynh-Dinh, Indik, Inta, Intini, Isa,
  Isac, Isi, Iyer, Izumi, Jacqmin, Jani, Jaranowski, Jawahar,
  Jim\'enez-Forteza, Johnson, Johnson-McDaniel, Jones, Jones, Jonker, Ju,
  Junker, Kalaghatgi, Kalogera, Kamai, Kandhasamy, Kang, Kanner, Kapadia,
  Karki, Karvinen, Kasprzack, Kastaun, Katolik, Katsavounidis, Katzman, Kaufer,
  Kawabe, K\'ef\'elian, Keitel, Kemball, Kennedy, Kent, Key, Khalili, Khan,
  Khan, Khan, Khazanov, Kijbunchoo, Kim, Kim, Kim, Kim, Kim, Kim, Kimbrell,
  King, King, Kinley-Hanlon, Kirchhoff, Kissel, Kleybolte, Klimenko, Knowles,
  Koch, Koehlenbeck, Koley, Kondrashov, Kontos, Korobko, Korth, Kowalska,
  Kozak, Kr\"amer, Kringel, Krishnan, Kr\'olak, Kuehn, Kumar, Kumar, Kumar,
  Kuo, Kutynia, Kwang, Lackey, Lai, Landry, Lang, Lange, Lantz, Lanza, Larson,
  Lartaux-Vollard, Lasky, Laxen, Lazzarini, Lazzaro, Leaci, Leavey, Lee, Lee,
  Lee, Lee, Lee, Lehmann, Lenon, Leon, Leonardi, Leroy, Letendre, Levin, Li,
  Linker, Littenberg, Liu, Liu, Lo, Lockerbie, London, Lord, Lorenzini,
  Loriette, Lormand, Losurdo, Lough, Lousto, Lovelace, L\"uck, Lumaca,
  Lundgren, Lynch, Ma, Macas, Macfoy, Machenschalk, MacInnis, Macleod, Maga\~na
  Hernandez, Maga\~na Sandoval, Maga\~na Zertuche, Magee, Majorana, Maksimovic,
  Man, Mandic, Mangano, Mansell, Manske, Mantovani, Marchesoni, Marion,
  M\'arka, M\'arka, Markakis, Markosyan, Markowitz, Maros, Marquina, Marsh,
  Martelli, Martellini, Martin, Martin, Martynov, Marx, Mason, Massera,
  Masserot, Massinger, Masso-Reid, Mastrogiovanni, Matas, Matichard, Matone,
  Mavalvala, Mazumder, McCarthy, McClelland, McCormick, McCuller, McGuire,
  McIntyre, McIver, McManus, McNeill, McRae, McWilliams, Meacher, Meadors,
  Mehmet, Meidam, Mejuto-Villa, Melatos, Mendell, Mercer, Merilh, Merzougui,
  Meshkov, Messenger, Messick, Metzdorff, Meyers, Miao, Michel, Middleton,
  Mikhailov, Milano, Miller, Miller, Miller, Millhouse, Milovich-Goff,
  Minazzoli, Minenkov, Ming, Mishra, Mitra, Mitrofanov, Mitselmakher,
  Mittleman, Moffa, Moggi, Mogushi, Mohan, Mohapatra, Molina, Montani, Moore,
  Moraru, Moreno, Morisaki, Morriss, Mours, Mow-Lowry, Mueller, Muir,
  Mukherjee, Mukherjee, Mukherjee, Mukund, Mullavey, Munch, Mu\~niz, Muratore,
  Murray, Nagar, Napier, Nardecchia, Naticchioni, Nayak, Neilson, Nelemans,
  Nelson, Nery, Neunzert, Nevin, Newport, Newton, Ng, Nguyen, Nguyen, Nichols,
  Nielsen, Nissanke, Nitz, Noack, Nocera, Nolting, North, Nuttall, Oberling,
  O'Dea, Ogin, Oh, Oh, Ohme, Okada, Oliver, Oppermann, Oram, O'Reilly,
  Ormiston, Ortega, O'Shaughnessy, Ossokine, Ottaway, Overmier, Owen, Pace,
  Page, Page, Pai, Pai, Palamos, Palashov, Palomba, Pal-Singh, Pan, Pan, Pang,
  Pang, Pankow, Pannarale, Pant, Paoletti, Paoli, Papa, Parida, Parker,
  Pascucci, Pasqualetti, Passaquieti, Passuello, Patil, Patricelli, Pearlstone,
  Pedraza, Pedurand, Pekowsky, Pele, Penn, Perez, Perreca, Perri, Pfeiffer,
  Phelps, Piccinni, Pichot, Piergiovanni, Pierro, Pillant, Pinard, Pinto,
  Pirello, Pitkin, Poe, Poggiani, Popolizio, Porter, Post, Powell, Prasad,
  Pratt, Pratten, Predoi, Prestegard, Prijatelj, Principe, Privitera, Prix,
  Prodi, Prokhorov, Puncken, Punturo, Puppo, P\"urrer, Qi, Quetschke, Quintero,
  Quitzow-James, Raab, Rabeling, Radkins, Raffai, Raja, Rajan, Rajbhandari,
  Rakhmanov, Ramirez, Ramos-Buades, Rapagnani, Raymond, Razzano, Read,
  Regimbau, Rei, Reid, Reitze, Ren, Reyes, Ricci, Ricker, Rieger, Riles, Rizzo,
  Robertson, Robie, Robinet, Rocchi, Rolland, Rollins, Roma, Romano, Romano,
  Romel, Romie, Rosi\ifmmode~\acute{n}\else \'{n}\fi{}ska, Ross, Rowan,
  R\"udiger, Ruggi, Rutins, Ryan, Sachdev, Sadecki, Sadeghian, Sakellariadou,
  Salconi, Saleem, Salemi, Samajdar, Sammut, Sampson, Sanchez, Sanchez,
  Sanchis-Gual, Sandberg, Sanders, Sassolas, Sathyaprakash, Saulson, Sauter,
  Savage, Sawadsky, Schale, Scheel, Scheuer, Schmidt, Schmidt, Schnabel,
  Schofield, Sch\"onbeck, Schreiber, Schuette, Schulte, Schutz, Schwalbe,
  Scott, Scott, Seidel, Sellers, Sengupta, Sentenac, Sequino, Sergeev,
  Shaddock, Shaffer, Shah, Shahriar, Shaner, Shao, Shapiro, Shawhan, Sheperd,
  Shoemaker, Shoemaker, Siellez, Siemens, Sieniawska, Sigg, Silva, Singer,
  Singh, Singhal, Sintes, Slagmolen, Smith, Smith, Smith, Somala, Son,
  Sonnenberg, Sorazu, Sorrentino, Souradeep, Spencer, Srivastava, Staats,
  Staley, Steinke, Steinlechner, Steinlechner, Steinmeyer, Stevenson, Stone,
  Stops, Strain, Stratta, Strigin, Strunk, Sturani, Stuver, Summerscales, Sun,
  Sunil, Suresh, Sutton, Swinkels, Szczepa\ifmmode~\acute{n}\else
  \'{n}\fi{}czyk, Tacca, Tait, Talbot, Talukder, Tanner, T\'apai, Taracchini,
  Tasson, Taylor, Taylor, Tewari, Theeg, Thies, Thomas, Thomas, Thomas, Thorne,
  Thorne, Thrane, Tiwari, Tiwari, Tokmakov, Toland, Tonelli, Tornasi,
  Torres-Forn\'e, Torrie, T\"oyr\"a, Travasso, Traylor, Trinastic, Tringali,
  Trozzo, Tsang, Tse, Tso, Tsukada, Tsuna, Tuyenbayev, Ueno, Ugolini,
  Unnikrishnan, Urban, Usman, Vahlbruch, Vajente, Valdes, Vallisneri, van
  Bakel, van Beuzekom, van~den Brand, Van Den~Broeck, Vander-Hyde, van~der
  Schaaf, van Heijningen, van Veggel, Vardaro, Varma, Vass, Vas\'uth, Vecchio,
  Vedovato, Veitch, Veitch, Venkateswara, Venugopalan, Verkindt, Vetrano,
  Vicer\'e, Viets, Vinciguerra, Vine, Vinet, Vitale, Vo, Vocca, Vorvick,
  Vyatchanin, Wade, Wade, Wade, Walet, Walker, Wallace, Walsh, Wang, Wang,
  Wang, Wang, Wang, Ward, Warner, Was, Watchi, Weaver, Wei, Weinert, Weinstein,
  Weiss, Wen, Wessel, We\ss{}els, Westerweck, Westphal, Wette, Whelan,
  Whitcomb, Whiting, Whittle, Wilken, Williams, Williams, Williamson, Willis,
  Willke, Wimmer, Winkler, Wipf, Wittel, Woan, Woehler, Wofford, Wong, Worden,
  Wright, Wu, Wysocki, Xiao, Yamamoto, Yancey, Yang, Yap, Yazback, Yu, Yu,
  Yvert, Zadro\ifmmode~\dot{z}\else \.{z}\fi{}ny, Zanolin, Zelenova, Zendri,
  Zevin, Zhang, Zhang, Zhang, Zhang, Zhao, Zhou, Zhou, Zhu, Zhu, Zimmerman,
  Zucker, \& Zweizig}]{ligo_ns}
Abbott, B.~P., {et~al.} 2017, Phys. Rev. Lett., 119, 161101

\bibitem[{Accadia {et~al.}(2012)Accadia, Acernese, Alshourbagy, Amico,
  Antonucci, Aoudia, Arnaud, Arnault, Arun, Astone, Avino, Babusci, Ballardin,
  Barone, Barrand, Barsotti, Barsuglia, Basti, Bauer, Beauville, Bebronne,
  Bejger, Beker, Bellachia, Belletoile, Beney, Bernardini, Bigotta, Bilhaut,
  Birindelli, Bitossi, Bizouard, Blom, Boccara, Boget, Bondu, Bonelli, Bonnand,
  Boschi, Bosi, Bouedo, Bouhou, Bozzi, Bracci, Braccini, Bradaschia, Branchesi,
  Briant, Brillet, Brisson, Brocco, Bulik, Bulten, Buskulic, Buy, Cagnoli,
  Calamai, Calloni, Campagna, Canuel, Carbognani, Carbone, Cavalier, Cavalieri,
  Cecchi, Cella, Cesarini, Chassande-Mottin, Chatterji, Chiche, Chincarini,
  Chiummo, Christensen, Clapson, Cleva, Coccia, Cohadon, Colacino, Colas,
  Colla, Colombini, Conforto, Corsi, Cortese, Cottone, Coulon, Cuoco,
  D'Antonio, Daguin, Dari, Dattilo, David, Davier, Day, Debreczeni, Carolis,
  Dehamme, Fabbro, Pozzo, del Prete, Derome, Rosa, DeSalvo, Dialinas, Fiore,
  Lieto, Emilio, Virgilio, Dietz, Doets, Dominici, Dominjon, Drago, Drezen,
  Dujardin, Dulach, Eder, Eleuteri, Enard, Evans, Fabbroni, Fafone, Fang,
  Ferrante, Fidecaro, Fiori, Flaminio, Forest, Forte, Fournier, Fournier,
  Franc, Francois, Frasca, Frasconi, Freise, Gaddi, Galimberti, Gammaitoni,
  Ganau, Garnier, Garufi, Gáspár, Gemme, Genin, Gennai, Gennaro, Giacobone,
  Giazotto, Giordano, Giordano, Girard, Gouaty, Grado, Granata, Granata, Grave,
  Greverie, Groenstege, Guidi, Hamdani, Hayau, Hebri, Heidmann, Heitmann,
  Hello, Hemming, Hennes, Hermel, Heusse, Holloway, Huet, Iannarelli,
  Jaranowski, Jehanno, Journet, Karkar, Ketel, Voet, Kovalik, Kowalska,
  Kreckelbergh, Krolak, Lacotte, Lagrange, Penna, Laval, Marec, Leroy,
  Letendre, Li, Lieunard, Liguori, Lodygensky, Lopez, Lorenzini, Loriette,
  Losurdo, Loupias, Mackowski, Maiani, Majorana, Magazzù, Maksimovic,
  Malvezzi, Man, Mancini, Mansoux, Mantovani, Marchesoni, Marion, Marin,
  Marque, Martelli, Masserot, Massonnet, Matone, Matone, Mazzoni, Menzinger,
  Michel, Milano, Minenkov, Mitra, Mohan, Montorio, Morand, Moreau, Moreau,
  Morgado, Morgia, Mosca, Moscatelli, Mours, Mugnier, Mul, Naticchioni, Neri,
  Nocera, Pacaud, Pagliaroli, Pai, Palladino, Palomba, Paoletti, Paoletti,
  Paoli, Pardi, Parguez, Parisi, Pasqualetti, Passaquieti, Passuello,
  Perciballi, Perniola, Persichetti, Petit, Pichot, Piergiovanni, Pietka,
  Pignard, Pinard, Poggiani, Popolizio, Pradier, Prato, Prodi, Punturo, Puppo,
  Qipiani, Rabaste, Rabeling, Rácz, Raffaelli, Rapagnani, Rapisarda, Re,
  Reboux, Regimbau, Reita, Remilleux, Ricci, Ricciardi, Richard, Ripepe,
  Robinet, Rocchi, Rolland, Romano, Rosińska, Roudier, Ruggi, Russo, Salconi,
  Sannibale, Sassolas, Sentenac, Solimeno, Sottile, Sperandio, Stanga, Sturani,
  Swinkels, Tacca, Taddei, Taffarello, Tarallo, Tissot, Toncelli, Tonelli,
  Torre, Tournefier, Travasso, Tremola, Turri, Vajente, van~den Brand, Broeck,
  van~der Putten, Vasuth, Vavoulidis, Vedovato, Verkindt, Vetrano, Véziant,
  Viceré, Vinet, Vilalte, Vitale, Vocca, Ward, Was, Yamamoto, Yvert, Zendri,
  \& Zhang}]{virgo}
Accadia, T., {et~al.} 2012, Journal of Instrumentation, 7, P03012

\bibitem[{{Barnes} {et~al.}(2016){Barnes}, {Kasen}, {Wu}, \&
  {Mart{\'{\i}}nez-Pinedo}}]{barnes}
{Barnes}, J., {Kasen}, D., {Wu}, M.-R., \& {Mart{\'{\i}}nez-Pinedo}, G. 2016,
  \apj, 829, 110

\bibitem[{{Beletic} {et~al.}(2008){Beletic}, {Blank}, {Gulbransen}, {Lee},
  {Loose}, {Piquette}, {Sprafke}, {Tennant}, {Zandian}, \& {Zino}}]{beletic}
{Beletic}, J.~W., {et~al.} 2008, in \procspie, Vol. 7021, High Energy, Optical,
  and Infrared Detectors for Astronomy III, 70210H

\bibitem[{{Bellm} \& {Kulkarni}(2017)}]{ztf}
{Bellm}, E., \& {Kulkarni}, S. 2017, Nature Astronomy, 1, 0071

\bibitem[{{Benford} {et~al.}(2008){Benford}, {Lauer}, \&
  {Mott}}]{BenfordLauerMott2008}
{Benford}, D.~J., {Lauer}, T.~R., \& {Mott}, D.~B. 2008, in \procspie, Vol.
  7021, High Energy, Optical, and Infrared Detectors for Astronomy III, 70211V

\bibitem[{{Bersten} {et~al.}(2018){Bersten}, {Folatelli}, {Garc{\'{\i}}a}, {van
  Dyk}, {Benvenuto}, {Orellana}, {Buso}, {S{\'a}nchez}, {Tanaka}, {Maeda},
  {Filippenko}, {Zheng}, {Brink}, {Cenko}, {de Jaeger}, {Kumar}, {Moriya},
  {Nomoto}, {Perley}, {Shivvers}, \& {Smith}}]{2016gkg_nature}
{Bersten}, M.~C., {et~al.} 2018, \nat, 554, 497

\bibitem[{{Bertin} \& {Arnouts}(1996)}]{sextractor}
{Bertin}, E., \& {Arnouts}, S. 1996, \aaps, 117, 393

\bibitem[{{Carnall} {et~al.}(2015){Carnall}, {Shanks}, {Chehade}, {Fumagalli},
  {Rauch}, {Irwin}, {Gonzalez-Solares}, {Findlay}, \&
  {Metcalfe}}]{carnall_atlas}
{Carnall}, A.~C., {et~al.} 2015, \mnras, 451, L16

\bibitem[{{Chapman} {et~al.}(1990){Chapman}, {Beard}, {Mountain}, {Pettie}, \&
  {Pickup}}]{chapman1990}
{Chapman}, R., {Beard}, S., {Mountain}, M., {Pettie}, D., \& {Pickup}, A. 1990,
  in \procspie, Vol. 1235, Instrumentation in Astronomy VII, ed. D.~L.
  {Crawford}, 34--42

\bibitem[{{Chen} {et~al.}(2017){Chen}, {Essick}, {Vitale}, {Holz}, \&
  {Katsavounidis}}]{chen_holz_15}
{Chen}, H.-Y., {Essick}, R., {Vitale}, S., {Holz}, D.~E., \& {Katsavounidis},
  E. 2017, \apj, 835, 31

\bibitem[{{Cowperthwaite} {et~al.}(2017{\natexlab{a}}){Cowperthwaite},
  {Berger}, {Rest}, {Chornock}, {Scolnic}, {Williams}, {Fong}, {Drout},
  {Foley}, {Margutti}, {Lunnan}, {Metzger}, \&
  {Quataert}}]{cowperthwaite_survey}
{Cowperthwaite}, P.~S., {et~al.} 2017{\natexlab{a}}, ArXiv e-prints

\bibitem[{{Cowperthwaite} {et~al.}(2017{\natexlab{b}}){Cowperthwaite},
  {Berger}, {Villar}, {Metzger}, {Nicholl}, {Chornock}, {Blanchard}, {Fong},
  {Margutti}, {Soares-Santos}, {Alexander}, {Allam}, {Annis}, {Brout}, {Brown},
  {Butler}, {Chen}, {Diehl}, {Doctor}, {Drout}, {Eftekhari}, {Farr}, {Finley},
  {Foley}, {Frieman}, {Fryer}, {Garc{\'{\i}}a-Bellido}, {Gill}, {Guillochon},
  {Herner}, {Holz}, {Kasen}, {Kessler}, {Marriner}, {Matheson}, {Neilsen},
  {Quataert}, {Palmese}, {Rest}, {Sako}, {Scolnic}, {Smith}, {Tucker},
  {Williams}, {Balbinot}, {Carlin}, {Cook}, {Durret}, {Li}, {Lopes}, {Louren{\c
  c}o}, {Marshall}, {Medina}, {Muir}, {Mu{\~n}oz}, {Sauseda}, {Schlegel},
  {Secco}, {Vivas}, {Wester}, {Zenteno}, {Zhang}, {Abbott}, {Banerji},
  {Bechtol}, {Benoit-L{\'e}vy}, {Bertin}, {Buckley-Geer}, {Burke}, {Capozzi},
  {Carnero Rosell}, {Carrasco Kind}, {Castander}, {Crocce}, {Cunha},
  {D'Andrea}, {da Costa}, {Davis}, {DePoy}, {Desai}, {Dietrich},
  {Drlica-Wagner}, {Eifler}, {Evrard}, {Fernandez}, {Flaugher}, {Fosalba},
  {Gaztanaga}, {Gerdes}, {Giannantonio}, {Goldstein}, {Gruen}, {Gruendl},
  {Gutierrez}, {Honscheid}, {Jain}, {James}, {Jeltema}, {Johnson}, {Johnson},
  {Kent}, {Krause}, {Kron}, {Kuehn}, {Nuropatkin}, {Lahav}, {Lima}, {Lin},
  {Maia}, {March}, {Martini}, {McMahon}, {Menanteau}, {Miller}, {Miquel},
  {Mohr}, {Neilsen}, {Nichol}, {Ogando}, {Plazas}, {Roe}, {Romer}, {Roodman},
  {Rykoff}, {Sanchez}, {Scarpine}, {Schindler}, {Schubnell}, {Sevilla-Noarbe},
  {Smith}, {Smith}, {Sobreira}, {Suchyta}, {Swanson}, {Tarle}, {Thomas},
  {Thomas}, {Troxel}, {Vikram}, {Walker}, {Wechsler}, {Weller}, {Yanny}, \&
  {Zuntz}}]{cowperthwaite_ns}
---. 2017{\natexlab{b}}, \apjl, 848, L17

\bibitem[{{de Val-Borro} {et~al.}(2016){de Val-Borro}, {Bakos}, {Brahm},
  {Hartman}, {Espinoza}, {Penev}, {Ciceri}, {Jord{\'a}n}, {Bhatti}, {Csubry},
  {Bayliss}, {Bento}, {Zhou}, {Rabus}, {Mancini}, {Henning}, {Schmidt}, {Tan},
  {Tinney}, {Wright}, {Kedziora-Chudczer}, {Bailey}, {Suc}, {Durkan},
  {L{\'a}z{\'a}r}, {Papp}, \& {S{\'a}ri}}]{deValBorro}
{de Val-Borro}, M., {et~al.} 2016, \aj, 152, 161

\bibitem[{{Drout} {et~al.}(2017){Drout}, {Piro}, {Shappee}, {Kilpatrick},
  {Simon}, {Contreras}, {Coulter}, {Foley}, {Siebert}, {Morrell}, {Boutsia},
  {Di Mille}, {Holoien}, {Kasen}, {Kollmeier}, {Madore}, {Monson},
  {Murguia-Berthier}, {Pan}, {Prochaska}, {Ramirez-Ruiz}, {Rest}, {Adams},
  {Alatalo}, {Ba{\~n}ados}, {Baughman}, {Beers}, {Bernstein}, {Bitsakis},
  {Campillay}, {Hansen}, {Higgs}, {Ji}, {Maravelias}, {Marshall}, {Moni Bidin},
  {Prieto}, {Rasmussen}, {Rojas-Bravo}, {Strom}, {Ulloa},
  {Vargas-Gonz{\'a}lez}, {Wan}, \& {Whitten}}]{drout}
{Drout}, M.~R., {et~al.} 2017, ArXiv e-prints

\bibitem[{{Fowler} \& {Gatley}(1990)}]{fowler}
{Fowler}, A.~M., \& {Gatley}, I. 1990, \apjl, 353, L33

\bibitem[{{Goobar} {et~al.}(2017){Goobar}, {Amanullah}, {Kulkarni}, {Nugent},
  {Johansson}, {Steidel}, {Law}, {M{\"o}rtsell}, {Quimby}, {Blagorodnova},
  {Brandeker}, {Cao}, {Cooray}, {Ferretti}, {Fremling}, {Hangard}, {Kasliwal},
  {Kupfer}, {Lunnan}, {Masci}, {Miller}, {Nayyeri}, {Neill}, {Ofek},
  {Papadogiannakis}, {Petrushevska}, {Ravi}, {Sollerman}, {Sullivan}, {Taddia},
  {Walters}, {Wilson}, {Yan}, \& {Yaron}}]{goobar17}
{Goobar}, A., {et~al.} 2017, Science, 356, 291

\bibitem[{{Kasen} {et~al.}(2013){Kasen}, {Badnell}, \& {Barnes}}]{kasen}
{Kasen}, D., {Badnell}, N.~R., \& {Barnes}, J. 2013, \apj, 774, 25

\bibitem[{{Kasen} {et~al.}(2015){Kasen}, {Fern{\'a}ndez}, \&
  {Metzger}}]{kasen2015}
{Kasen}, D., {Fern{\'a}ndez}, R., \& {Metzger}, B.~D. 2015, \mnras, 450, 1777

\bibitem[{{Kasen} {et~al.}(2017){Kasen}, {Metzger}, {Barnes}, {Quataert}, \&
  {Ramirez-Ruiz}}]{kasen_nature}
{Kasen}, D., {Metzger}, B., {Barnes}, J., {Quataert}, E., \& {Ramirez-Ruiz}, E.
  2017, ArXiv e-prints

\bibitem[{{Kasliwal} {et~al.}(2017){Kasliwal}, {Nakar}, {Singer}, {Kaplan},
  {Cook}, {Van Sistine}, {Lau}, {Fremling}, {Gottlieb}, {Jencson}, {Adams},
  {Feindt}, {Hotokezaka}, {Ghosh}, {Perley}, {Yu}, {Piran}, {Allison},
  {Anupama}, {Balasubramanian}, {Bannister}, {Bally}, {Barnes}, {Barway},
  {Bellm}, {Bhalerao}, {Bhattacharya}, {Blagorodnova}, {Bloom}, {Brady},
  {Cannella}, {Chatterjee}, {Cenko}, {Cobb}, {Copperwheat}, {Corsi}, {De},
  {Dobie}, {Emery}, {Evans}, {Fox}, {Frail}, {Frohmaier}, {Goobar}, {Hallinan},
  {Harrison}, {Helou}, {Hinderer}, {Ho}, {Horesh}, {Ip}, {Itoh}, {Kasen},
  {Kim}, {Kuin}, {Kupfer}, {Lynch}, {Madsen}, {Mazzali}, {Miller}, {Mooley},
  {Murphy}, {Ngeow}, {Nichols}, {Nissanke}, {Nugent}, {Ofek}, {Qi}, {Quimby},
  {Rosswog}, {Rusu}, {Sadler}, {Schmidt}, {Sollerman}, {Steele}, {Williamson},
  {Xu}, {Yan}, {Yatsu}, {Zhang}, \& {Zhao}}]{kasliwal_nsns}
{Kasliwal}, M.~M., {et~al.} 2017, ArXiv e-prints

\bibitem[{{Law} {et~al.}(2009){Law}, {Kulkarni}, {Dekany}, {Ofek}, {Quimby},
  {Nugent}, {Surace}, {Grillmair}, {Bloom}, {Kasliwal}, {Bildsten}, {Brown},
  {Cenko}, {Ciardi}, {Croner}, {Djorgovski}, {van Eyken}, {Filippenko}, {Fox},
  {Gal-Yam}, {Hale}, {Hamam}, {Helou}, {Henning}, {Howell}, {Jacobsen},
  {Laher}, {Mattingly}, {McKenna}, {Pickles}, {Poznanski}, {Rahmer}, {Rau},
  {Rosing}, {Shara}, {Smith}, {Starr}, {Sullivan}, {Velur}, {Walters}, \&
  {Zolkower}}]{iptf}
{Law}, N.~M., {et~al.} 2009, \pasp, 121, 1395

\bibitem[{{Li} \& {Paczy{\'n}ski}(1998)}]{lipaczynski}
{Li}, L.-X., \& {Paczy{\'n}ski}, B. 1998, \apjl, 507, L59

\bibitem[{{LSST Science Collaboration} {et~al.}(2009){LSST Science
  Collaboration}, {Abell}, {Allison}, {Anderson}, {Andrew}, {Angel}, {Armus},
  {Arnett}, {Asztalos}, {Axelrod}, \& et~al.}]{lsst}
{LSST Science Collaboration} {et~al.} 2009, ArXiv e-prints

\bibitem[{{Margutti} {et~al.}(2017){Margutti}, {Berger}, {Fong}, {Guidorzi},
  {Alexander}, {Metzger}, {Blanchard}, {Cowperthwaite}, {Chornock},
  {Eftekhari}, {Nicholl}, {Villar}, {Williams}, {Annis}, {Brown}, {Chen},
  {Doctor}, {Frieman}, {Holz}, {Sako}, \& {Soares-Santos}}]{offax_jet}
{Margutti}, R., {et~al.} 2017, \apjl, 848, L20

\bibitem[{{Menanteau} {et~al.}(2012){Menanteau}, {Hughes}, {Sif{\'o}n},
  {Hilton}, {Gonz{\'a}lez}, {Infante}, {Barrientos}, {Baker}, {Bond}, {Das},
  {Devlin}, {Dunkley}, {Hajian}, {Hincks}, {Kosowsky}, {Marsden}, {Marriage},
  {Moodley}, {Niemack}, {Nolta}, {Page}, {Reese}, {Sehgal}, {Sievers},
  {Spergel}, {Staggs}, \& {Wollack}}]{menanteau2012}
{Menanteau}, F., {et~al.} 2012, \apj, 748, 7

\bibitem[{{Metzger} \& {Fern{\'a}ndez}(2014)}]{metzger2015}
{Metzger}, B.~D., \& {Fern{\'a}ndez}, R. 2014, \mnras, 441, 3444

\bibitem[{{Metzger} {et~al.}(2010){Metzger}, {Mart{\'{\i}}nez-Pinedo},
  {Darbha}, {Quataert}, {Arcones}, {Kasen}, {Thomas}, {Nugent}, {Panov}, \&
  {Zinner}}]{metzger}
{Metzger}, B.~D., {et~al.} 2010, \mnras, 406, 2650

\bibitem[{{Mooley} {et~al.}(2018){Mooley}, {Nakar}, {Hotokezaka}, {Hallinan},
  {Corsi}, {Frail}, {Horesh}, {Murphy}, {Lenc}, {Kaplan}, {de}, {Dobie},
  {Chandra}, {Deller}, {Gottlieb}, {Kasliwal}, {Kulkarni}, {Myers}, {Nissanke},
  {Piran}, {Lynch}, {Bhalerao}, {Bourke}, {Bannister}, \& {Singer}}]{mooley}
{Mooley}, K.~P., {et~al.} 2018, \nat, 554, 207

\bibitem[{{Nakar} \& {Sari}(2012)}]{nakar2012}
{Nakar}, E., \& {Sari}, R. 2012, \apj, 747, 88

\bibitem[{{P{\'a}l}(2012)}]{fitsh}
{P{\'a}l}, A. 2012, \mnras, 421, 1825

\bibitem[{{Persson} {et~al.}(2013){Persson}, {Murphy}, {Smee}, {Birk},
  {Monson}, {Uomoto}, {Koch}, {Shectman}, {Barkhouser}, {Orndorff}, {Hammond},
  {Harding}, {Scharfstein}, {Kelson}, {Marshall}, \& {McCarthy}}]{fourstar}
{Persson}, S.~E., {et~al.} 2013, \pasp, 125, 654

\bibitem[{{Piro} \& {Kollmeier}(2017)}]{piro}
{Piro}, A.~L., \& {Kollmeier}, J.~A. 2017, ArXiv e-prints

\bibitem[{{Skrutskie} {et~al.}(2006){Skrutskie}, {Cutri}, {Stiening},
  {Weinberg}, {Schneider}, {Carpenter}, {Beichman}, {Capps}, {Chester},
  {Elias}, {Huchra}, {Liebert}, {Lonsdale}, {Monet}, {Price}, {Seitzer},
  {Jarrett}, {Kirkpatrick}, {Gizis}, {Howard}, {Evans}, {Fowler}, {Fullmer},
  {Hurt}, {Light}, {Kopan}, {Marsh}, {McCallon}, {Tam}, {Van Dyk}, \&
  {Wheelock}}]{2mass}
{Skrutskie}, M.~F., {et~al.} 2006, \aj, 131, 1163

\bibitem[{Sullivan(2015)}]{sullivan_thesis}
Sullivan, P.~W. 2015, Master's thesis, MIT, Cambridge, MA 02139

\bibitem[{{Sullivan} {et~al.}(2013){Sullivan}, {Croll}, \&
  {Simcoe}}]{sullivan_ap640c}
{Sullivan}, P.~W., {Croll}, B., \& {Simcoe}, R.~A. 2013, \pasp, 125, 1021

\bibitem[{{Sullivan} {et~al.}(2014){Sullivan}, {Croll}, \&
  {Simcoe}}]{sullivan_ap1121}
{Sullivan}, P.~W., {Croll}, B., \& {Simcoe}, R.~A. 2014, in \procspie, Vol.
  9154, High Energy, Optical, and Infrared Detectors for Astronomy VI, 91541F

\bibitem[{{Tokunaga} {et~al.}(2002){Tokunaga}, {Simons}, \& {Vacca}}]{tokunaga}
{Tokunaga}, A.~T., {Simons}, D.~A., \& {Vacca}, W.~D. 2002, \pasp, 114, 180

\bibitem[{{Villar} {et~al.}(2017){Villar}, {Guillochon}, {Berger}, {Metzger},
  {Cowperthwaite}, {Nicholl}, {Alexander}, {Blanchard}, {Chornock},
  {Eftekhari}, {Fong}, {Margutti}, \& {Williams}}]{villar}
{Villar}, V.~A., {et~al.} 2017, \apjl, 851, L21

\end{thebibliography}

\end{document}